\documentclass[final,5p,times,twocolumn]{elsarticle}

\usepackage{amssymb}
\usepackage{color}
\usepackage{ulem}
\usepackage{subcaption}
\journal{Aerospace Science and Technology(AESCTE)}

\begin{document}

\begin{frontmatter}



\title{
CFD Analysis on the Performance of a Coaxial Rotor \\ with Lift Offset at High Advance Ratios
}



\author[inst1]{Kaito Hayami}

\affiliation[inst1]{organization={Department of Mechanical Systems Engineering, Tokyo University of Agriculture and Technology},%Department and Organization
            addressline={Naka-cho 2-24-16}, 
            city={Koganei},
            postcode={184-8588}, 
            state={Tokyo},
            country={Japan}}

\author[inst2]{Hideaki Sugawara}
\author[inst1]{Takumi Yumino}
\author[inst2]{Yasutada Tanabe}
\author[inst1]{Masaharu Kameda\corref{cor1}}

\affiliation[inst2]{organization={The Aviation Technology Directorate, Japan Aerospace Exploration Agency},%Department and Organization
            addressline={Osawa 6-13-1}, 
            city={Mitaka},
            postcode={181-0015}, 
            state={Tokyo},
            country={Japan}}

\ead{kame@cc.tuat.ac.jp,}
\cortext[cor1]{Corresponding author. Tel.: +81-42-388-7075; fax: +81-42-388-7413.}

\begin{abstract}
The aerodynamic performance of an isolated coaxial rotor in forward flight is analyzed by a high-fidelity computational fluid dynamics (CFD) approach. 
The analysis focuses on the high-speed forward flight with an advance ratio of 0.5 or higher, which is the ratio of the forward speed to the rotor tip speed. 
The effect of the degree of the rolling moment on the rotor thrust, called lift offset, is studied in detail. 
The coaxial rotor model is a pair of contrarotating rotors, each rotor consisting of two {untwisted} blades with a radius of 1.016 m.
The pitch angle of the blades is controlled by both collective and cyclic as in a conventional {single main-rotor} helicopter. 
CFD analysis is performed using a flow solver based on the compressible Navier-Stokes equations with a Reynolds-averaged turbulence model.
Laminar/turbulent transition in the boundary layer is taken into account in the calculation. 
The rotor trim for target forces and moments is achieved using a gradient-based delta-form blade pitch angle adjusting technique in conjunction with CFD analysis. 
The reliability of the calculations is confirmed by comparison with published wind tunnel experiments and two comprehensive analyses. 
Applying the lift offset improves the lift-to-effective drag ratio (lift-drag ratio) and reduces thrust fluctuations.
However, in the case where the advance ratio exceeds 0.6, the lift-drag ratio drops significantly even if the lift offset is 0.3.
The thrust fluctuation also increases with such a high advance ratio.
Detailed analysis reveals that the degradation of aerodynamic performance and {vibratory aerodynamic loads} is closely related to the pitch angle control to compensate for the reduction in thrust on the retreating side due to the increased reverse flow region.
It is effective to reduce the collective and longitudinal cyclic pitch angles for the improvement of the aerodynamic performance of coaxial rotors with an appropriate lift offset.
\end{abstract}




\end{frontmatter}

%%\linenumbers

\section{Introduction}
\label{sec:sample1}

Advanced {rotorcraft has} been actively studied in recent years \cite{Ormiston2016, Enconniere2017}. The development of high-speed {rotorcraft} is a major issue in this field \cite{Yeo2019}. 
The  {single main-rotor} system employed in conventional helicopters is not suitable for high-speed flights such as 500 km/h, because the relative flow velocity into the rotor blade during rotation is substantially different at the advancing side from the retreating side. 
This velocity difference causes difficulty in the achievement of a rolling moment balance as well as an optimization of lift and drag, even though a complex swash-plate mechanism is employed in conventional helicopters to vary the pitch angle sinusoidally.

\begin{figure}
\centering
  \includegraphics[width=\columnwidth]{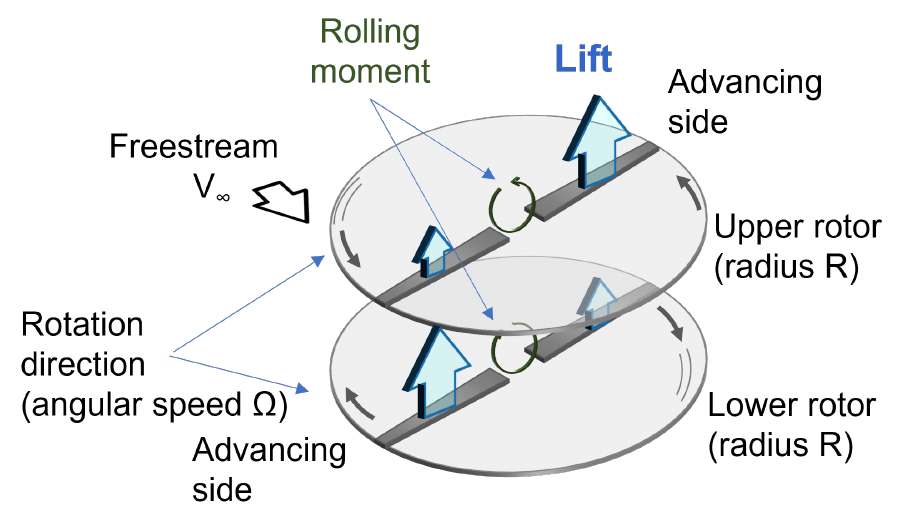}
\caption{An isolated coaxial rotor with lift offset.}
\label{fig:1}
\end{figure}

The coaxial rotor system is attractive for high-speed rotorcraft because it has the potential to overcome the drawbacks of the single rotor system. The system consists of a pair of contrarotating rotors as shown in Fig. \ref{fig:1}. It is not necessary to balance the rolling moments of each rotor if the net rolling moment of the system is balanced. This relaxation of the limitation for the rolling moment balance allows the rotor to generate more lift on the advancing side than on the retreating side. If it is not necessary to generate lift on the retreating side, the rotor lift-drag ratio is improved owing to the reduction of the large drag induced by relative reverse flow at the retreating side. 

The principle of operation mentioned above is called the “advancing blade concept (ABC)”, which was first proposed by Sikorsky aircraft in the 1960s. Then, the first demonstrator XH-59A was prototyped by the same company in the 1970s \cite{Ruddell1981}. 
After retiring the XH-59A due to its high levels of  {vibratory aerodynamic loads?} and fuel consumption, the subsequent series of demonstrator X2 Technology Demonstrator (X2TD) and S-97 RAIDER were developed by Sikorsky aircraft in this century \cite{Bagai2008, Walsh2011, Lorber2016}. 
Rapid growth in multirotor aircraft from the 2010s also shed light on the coaxial rotor system \cite{Zhu2020,Jamalhaddadi2022}. 

The aerodynamic performance of an isolated coaxial rotor without fuselage and pusher propeller is fundamental knowledge to design the rotorcraft with a coaxial rotor system. 
This performance is also useful to evaluate the feasibility of other concepts of high-speed rotorcraft such as winged compound helicopters \cite{Sugawara2019, Sugawara2021}.

The aerodynamic performance is characterized using the degree of the rolling moment called the lift offset, which is defined as 
\begin{equation}
    LO = \frac{\left|C_{Mx}^U\right| + \left|C_{Mx}^L\right|}{C_T^U + C_T^L} = \frac{r_T}{R},
\label{eq:LO}
\end{equation}
where $R$ and $r_T$ denote the rotor radius and the distance from the center of rotation to the rolling moment balancing position. 
$C_T$ and $C_{Mx}$ denote the thrust coefficient and the rolling moment coefficient, which are defined as
\begin{eqnarray}
    C_T & = & \frac{T}{\rho_\infty \pi R^2 \left(\Omega R\right)^2}, \label{eq:CT}\\
    C_{Mx} & = & \frac{M_x}{\rho_\infty \pi R^3 \left(\Omega R\right)^2}, \label{eq:CMx}
\end{eqnarray}
where $T$, $M_x$, $\rho_\infty$, $\Omega$, and $\pi$ are the rotor thrust, rotor rolling moment, air density in the  {freestream}, blade rotation angular velocity, and circle ratio, respectively. 
Superscripts $U$ and $L$ indicate the values for upper and lower rotors.  
Another parameter for the characterization is the advance ratio, which is defined as 
\begin{equation}
    \mu = \frac{V_{\infty} \cos i}{\Omega R},
    \label{eq:mu}
\end{equation}
where $V_\infty$ and $i$ denote the  {freestream velocity} and rotor disk plane incidence angle, respectively. 

Several numbers of literature are available as public information related to the coaxial rotors and lift offset technique. 
Yeo and Johnson \cite{Yeo2014} showed by the comprehensive analysis based on lifting line theory that the maximum lift capability of the coaxial rotor increases as lift offset increases. 
For example, the lift offset was 0.8 when the maximum lift was achieved at the advance ratio of 0.41. 
Subsequently, Cameron and Sirohi \cite{Cameron2016a} (detail in \cite{Cameron2016b}) conducted a wind tunnel experiment on the coaxial rotors where each rotor had a rotor diameter of 2.026 m and two blades.
In this experiment, the effect of lift offset on the aerodynamic performance of the coaxial rotors was investigated at the advance ratio $\mu$ from 0.21 to 0.52 with the lift offset $LO$ from 0 to 0.2. 
They clarified within their tested advance ratio that an appropriate lift offset $LO$ improved the aerodynamic performance such as the lift-drag ratio and reduced  {the vertical force of vibratory hub loads}. 
These tendencies were validated by the comprehensive analyses \cite{Feil2019, Feil2020, Ho2020a}.  

More recently, Ho and Yeo \citep{Ho2020b} analyzed the performance of an isolated coaxial rotor with a wide range of advance ratios ($0.3 \le \mu \le 0.65$) using the comprehensive analysis solver, the Rotorcraft Comprehensive Analysis System (RCAS). 
As a result, they suggested that the lift offset improved aerodynamic performance even under conditions of high advance ratio ($\mu$ = 0.65 (230 knots)). 
According to their analysis, a large lift offset ($LO$ = 0.3) significantly reduces the rotor shaft power, but slightly increases the drag of the rotor. 
This analysis shows the coaxial rotor system is feasible in high-speed flight, but there are still some points to be clarified. 
Ho and Yeo \cite{Ho2020b} considered structural deformation in their analysis. 
Flapping occurring at high advance ratios may cause the plane of rotation to tilt backward.
Therefore, it is not distinguished whether the decrease in rotor shaft power is due to the backward tilt of the rotary surface or purely due to the change in the aerodynamic performance of the rotor with the lift offset. 
Moreover, it is not confirmed by the wind tunnel tests whether the handling of vortex elements, which is the basis of comprehensive analysis, is appropriate under a high advance ratio \cite{Johnson2009}.  

Analysis based on computational fluid dynamics (CFD) has been also performed as a higher fidelity method than the comprehensive analysis. 
Passe et al. \cite{Passe2015} performed a CFD analysis of aerodynamic interference under forward flight conditions (0.15 (55 knots) $\le \mu \le$ 0.41 (150 knots)) using a coaxial rotor model based on the Sikorsky X2 Technology Demonstrator. They compared the CFD results with the results of the comprehensive analysis coupled with a vortex wake model. 
They showed that there were two types of aerodynamic interference that occurred in coaxial rotors, whose origins were different from each other. 
The low-frequency fluctuation component was generated when the wake generated by the upper rotor interfered with the lower rotor. 
This low-frequency phenomenon was captured with CFD and comprehensive analysis. 
The high-frequency thrust fluctuations were caused by aerodynamic interference due to the crossing of the upper and lower rotor blades. 
This high-frequency thrust fluctuation was captured only by CFD analysis, but not by the comprehensive analysis. 
Passe et al. \cite{Passe2015} concluded that a high-fidelity analysis such as CFD was required to accurately predict the aerodynamic characteristics of the coaxial rotor system. 

Subsequently, Jia and Lee \cite{Jia2020} analyzed the impulsive acoustic characteristics of a lift-offset coaxial rotor in high-speed forward flight (0.26 (100 knots) $\le \mu \le$ 0.59 (200 knots)) by a high-fidelity CFD/CSD (computational structural dynamics) loose-coupling simulation. 
They showed the lift-offset coaxial rotor model based on the XH-59A generated a dramatically higher impulsive sound-pressure level than the  {single rotor}, especially at high speed, as a consequence of mainly blade-crossover events. 
They assumed the lift offset to be 0.2 for all the calculations. Therefore, they did not show the correlation between this thrust fluctuation and the lift offset. 

In this study, we conduct a CFD analysis on the aerodynamic performance of an isolated coaxial rotor in forward flight. 
We mainly focus our attention on the performance at a high advance ratio over 0.5, even though we choose the range of advance ratio from 0.3 to 0.8 in the calculation. 
The advance ratio over 0.7 with a cruise speed of 250 knots (463 km/h) is the value expected for the advanced rotorcraft \cite{Johnson2009}, 
although it is not investigated in the aforementioned CFD studies. 
In the analysis, we clarify the correlation of lift offset with aerodynamic performance and thrust fluctuation by choosing a wide range of lift offsets. 

\section{Coaxial Rotor Model}
\label{sec:2}

\begin{figure}
\centering
\begin{tabular}{c}
\begin{minipage}{0.835\columnwidth}
    \includegraphics[width=\columnwidth]{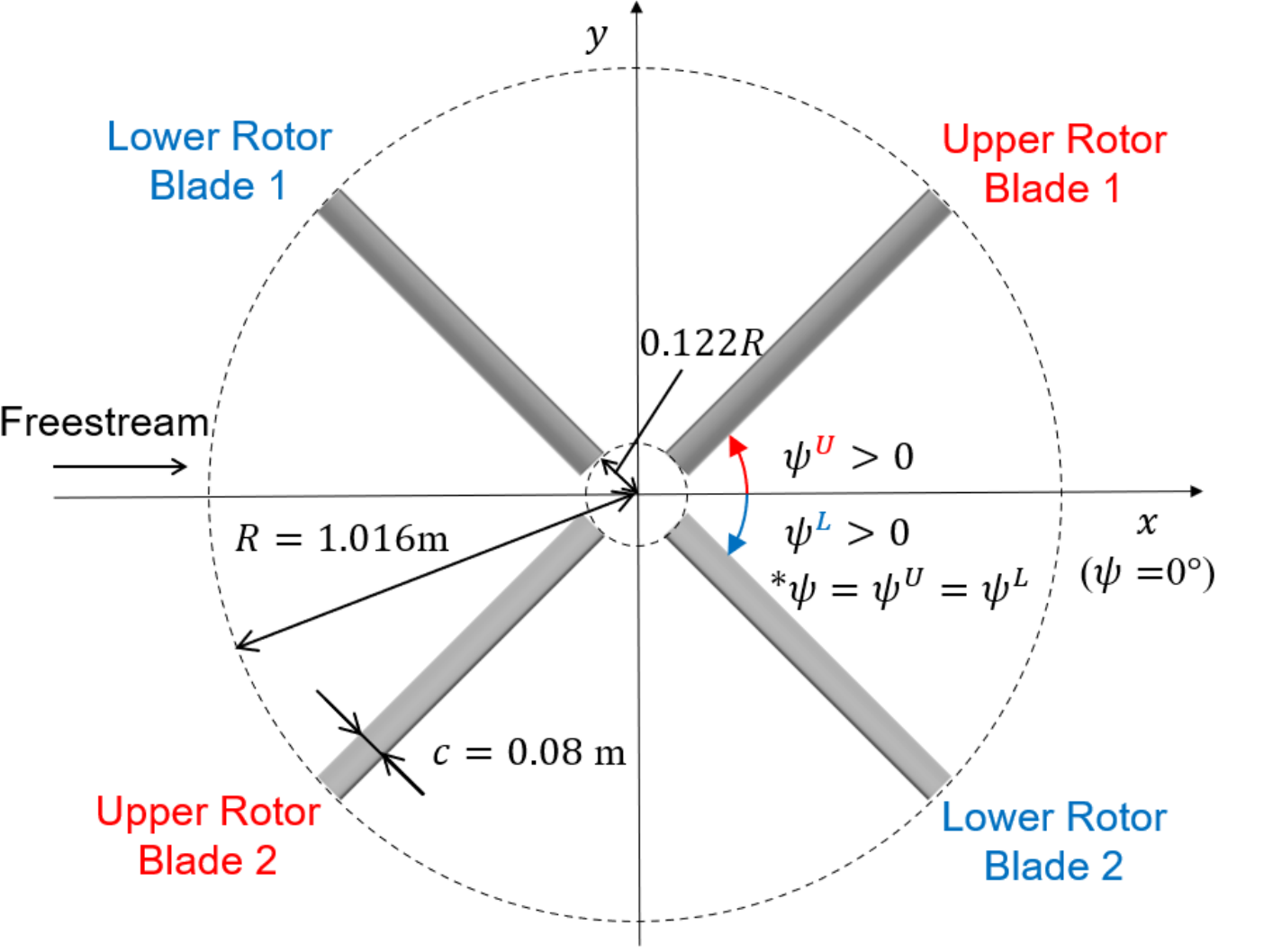}
    \subcaption{Top view}
    \label{fig:2_a}
\end{minipage}
\\
\begin{minipage}{0.8\columnwidth}
    \includegraphics[width=\columnwidth]{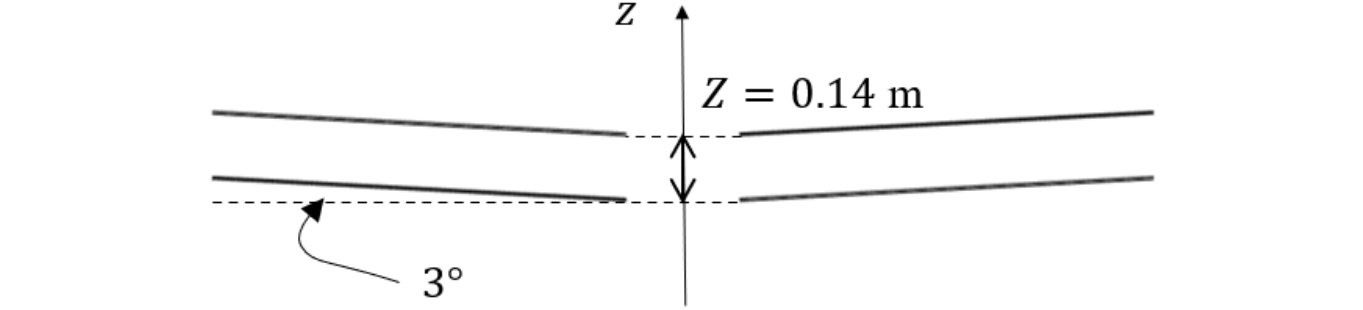}
    \subcaption{Side view}
    \label{fig:2_b}
\end{minipage}
\end{tabular}
\caption{Computational model.}
\label{fig:2}
\end{figure}

Figure \ref{fig:2} shows the computational model of the coaxial rotor system used in this study. 
It is a simplified system used in the wind-tunnel experiment by Cameron and Sirohi \cite{Cameron2016a} because its specifications and aerodynamic performance have been fully documented in published reports, including for high advance ratios up to 0.52 \cite{Cameron2016a, Cameron2016b, Feil2019, Ho2020a}. 
The system consists of a  {pair} of two-bladed hingeless rotors. 
The blade used is an untwisted VR-12  {\sout{rectangular}} airfoil. 
The rotor radius $R$ is 1.016 m, the root chord $c$ is 0.08 m, 
and the separation distance between the two rotors $Z$ is 0.14 m. 
Blade root-cut is set at $r = 0.122R$ in this calculation.
The solidity of the coaxial rotor system $\sigma$ is 0.100 where $\sigma = 4c/\pi R$. 
Each blade features a precone angle of 3$^\circ$. 

The rotational directions of the upper and lower rotors are counterclockwise and clockwise, respectively, when the rotor is viewed from above (see Fig. \ref{fig:2}). 
The azimuth angles of each rotor, $\Psi^U$ and $\Psi^L$, are defined as positive in their rotational directions with respect to the forward flight downstream direction. 
The angle of the blades of rotors (blade pitch $\theta$) is controlled by both collective and cyclic as in a conventional  {single main-rotor} helicopter. 
The blade pitch angle $\theta$ for each blade azimuth angle $\Psi$ is expressed as 
\begin{equation}
    \theta = \theta_0 + \theta_{\rm 1c} \cos \Psi + \theta_{\rm 1s} \sin \Psi, 
    \label{eq:theta}
\end{equation}
where the control angles $\theta_0$, $\theta_{\rm 1c}$, and $\theta_{\rm 1s}$ represent the collective pitch angle, the lateral cyclic pitch angle, and the longitudinal cyclic pitch angle, respectively. 
These angles are adjusted by the trim analysis based on the averaged rotor aerodynamic forces and moments. 
In this analysis, the blade crossing is assumed to occur at $\Psi = 0^{\circ}$, 90$^{\circ}$, 180$^{\circ}$, and 270$^{\circ}$.
The blade lead-lag motion and the blade elastic deformations are not considered under the assumption of a highly rigid blade. The flapping angles are set to 0$^{\circ}$. 
These assumptions allow us to understand the aerodynamic performance without the influence of structural deformations. 

\section{Numerical simulation}

\subsection{Flow solver}

A CFD flow solver specifically for rotorcraft, rFlow3D developed at JAXA \cite{Tanabe2009,Tanabe2010,Tanabe2012a,Tanabe2012b} is used to simulate the aerodynamic performance of the coaxial rotor in forward flight. 
rFlow3D is not just a CFD solver, but also covers trim analysis and elastic deformation analysis of blades. 
rFlow3D has been validated to give reliable solutions for many flows associated with rotorcraft.

The basic equations and numerical methods used in rFlow3D have been presented in detail in previous papers (e.g. \cite{Sugawara2019,Sugawara2021}).  
Only a brief description of the flow solver is addressed in this paper. 
rFlow3D employs a finite volume scheme for structured grids that can handle moving overlapping grids. 
The basic equations are the compressible Naiver-Stokes equations. Several turbulence models can be incorporated depending on the characteristics of the flow field. 
The coaxial rotor analyzed in this study does not have a Reynolds number large enough to assume that the flow is fully turbulent.
Therefore the transition model ($\gamma-Re_{\theta}$ model) \cite{Langtry2009} is employed to account for the laminar/turbulent transition in the boundary layer. 
A class of the Menter's shear stress transport (SST) turbulence model (so-called $k-\omega$ SST or SST-2003) \cite{Menter2003} is applied only to the region downstream of the transition point obtained by the transition model. 

\subsection{Trim analysis}

The trim analysis technique in conjunction with CFD analysis developed by Sugawara et al. \cite{Sugawara2022} is applied to the coaxial rotor system used in this study.
This technique is an extension of the trim analysis for single rotor systems \cite{Sugawara2019,Tanabe2012a}. 

The following six components are selected for trim analysis: the total rotor thrust of the coaxial rotor system $\sum C_T = C_T^U + C_T^L$, the rolling and pitching moments of the upper and lower rotor $C_{Mx}^U, C_{Mx}^L, C_{My}^U, C_{My}^L$, and the yawing moment (torque) balance $\sum C_{Mz} = C_{Mz}^U + C_{Mz}^L$. 
The moment coefficients $C_{My}$ and $C_{Mz}$ are defined as 
\begin{eqnarray}
    C_{My} & = & \frac{M_y}{\rho_\infty \pi R^3 \left(\Omega R\right)^2}, \label{eq:CMy} \\
        C_{Mz} & = & \frac{M_z}{\rho_\infty \pi R^3 \left(\Omega R\right)^2}, \label{eq:CMz}
\end{eqnarray}
where $M_y$ and $M_z$ are rotor pitching and yawing moments, respectively. 
These parameters are adjusted by the six parameters for determining blade pitch angles in Eq. (\ref{eq:theta}), which are $\theta_0^U$, $\theta_0^L$, $\theta_{\rm 1c}^U$, $\theta_{\rm 1c}^L$, $\theta_{\rm 1s}^U$ and $\theta_{\rm 1s}^L$. 
The sensitivity matrix of control elements $\theta_j$ to target variables $a_i$, $\left\{\partial a_i/\partial \theta_j\right\} (i,j =1-6)$, are numerically calculated using the blade element theory (BET) for a single blade \cite{Leishman2006}. 
Note that the sensitivity matrix calculation does not take into account the aerodynamic interference between the upper and lower rotors.
Although this interference might affect the target aerodynamic parameters, Sugawara et al. have confirmed that a smoothly converging solution is obtained without taking it into account \cite{Sugawara2022}. 

Each pitch angle parameter is updated by finding the difference between each aerodynamic coefficient obtained from CFD and its target value and multiplying the difference by the inverse of the sensitivity matrix.
The update is performed at every one revolution of the rotor. 

\subsection{Computational setup}

\begin{figure}
\centering
  \includegraphics[width=\columnwidth]{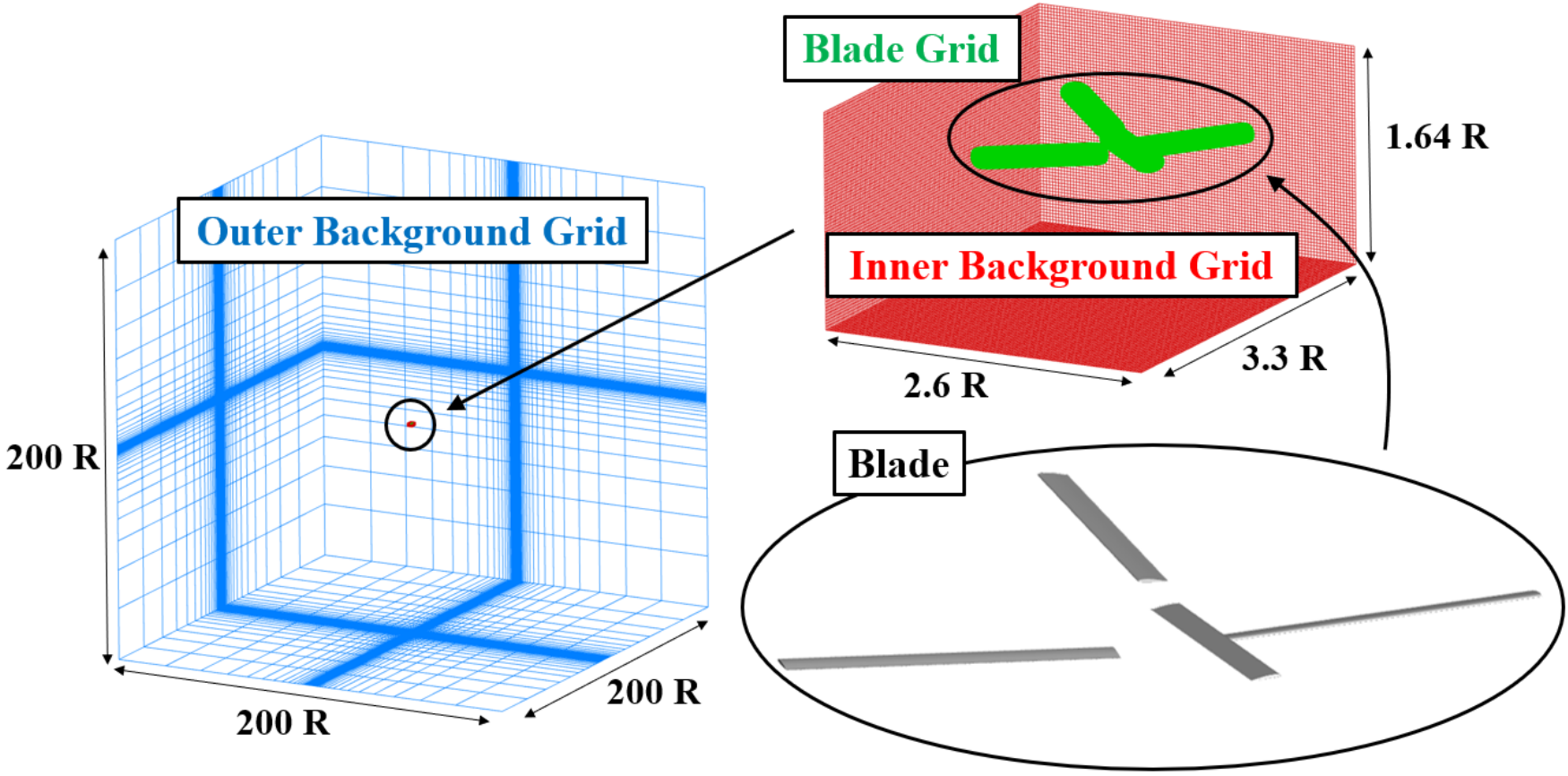}
\caption{Grid distributions of the coaxial rotor.}
\label{fig:3}
\end{figure}

The flow around the coaxial rotor without the rotor hub is analyzed using the computational grid shown in Fig. \ref{fig:3}. 
The total number of computational grid points is about 20.9 million, consisting of $125 \times 123 \times 65$ points on the blade grid, $279 \times 221 \times 139$ points on the inner background grid, and $211 \times 211 \times 169$ points on the outer background grid. 
The blade grid is a moving body-fitted grid, which means that the grid arrangement changes as the blade moves.
The computational domain of the outer background grid is set to $\pm 100R$ for each axis. 
The minimum grid spacing on the blade surface is set to be less than $3.0y^{+}$, where $y^+$ is the dimensionless length expressed using the friction velocity at the trailing edge of the tip of the rotor blade at the advance ratio $\mu = 0.2$. 
 {This minimum grid spacing is a reasonable value that adequately predicts drag using the SST turbulence model \cite{Smith2013}.}
Note that the boundary layer on the blades is not fully turbulent in this study, because the Reynolds number at 75\%$R$ is in the range between $1.2 \times 10^5$ and $6.6 \times 10^5$ at $\mu = 0.52$.
For all calculations, the pressure and temperature of the  {freestream} $p_{\infty}$ and $T_{\infty}$ are set to 101.3 kPa and 15$^{\circ}$C, respectively.
 {
For the calculation, one node of the Intel Xeon Gold 6240 processor (18 cores/36 threads, clock frequency 2.60 GHz) is used for parallel computation with 36 threads using OpenMP. 
The computation time per rotor revolution is 18 hours, and the quasi-steady solution is obtained after about 15 rotations.
}.

\subsection{Verification of the accuracy of calculations}

Prior to calculating with a high advance ratio of 0.5 or higher, the calculated aerodynamic performance is verified with wind tunnel test data obtained by Cameron and Sirohi \cite{Cameron2016a} and associated numerical data using two rotorcraft comprehensive analysis tools, the Comprehensive Analytical Model of Rotorcraft Aerodynamics and Dynamics (CAMRAD II)  {\cite{Johnson1994}} presented by Feil et al. \cite{Feil2019} and the Rotorcraft Comprehensive Analysis System (RCAS)  {\cite{Saberi2015}} presented by Ho and Yeo \cite{Ho2020a}.
Cameron and Sirohi \cite{Cameron2016a} provided extensive data on the aerodynamics performances of the single and coaxial rotors for a wide range of advance ratio  {($0.21 \le \mu \le 0.52$)} and lift offset ($LO = 0$--0.3). 
Very recently, Sugawara et al. \citep{Sugawara2022} have validated the numerical results with the CFD flow solver, rFlow3D, for coaxial rotor configurations with various lift offsets at an advance ratio of 0.53, comparing them with experiments and comprehensive analyses. 
The comparison reveals satisfactory agreement between their results and the other data. 
Therefore, the aerodynamic performance of the coaxial rotor for various advance ratios with zero lift offset is solely presented in this section. 

\begin{table}
\caption{Trim conditions for verification calculations}
\label{tbl:1}
    \centering
\small{
    \begin{tabular}{lcccc}
    \hline
        $\mu$ & 0.21 & 0.31 & 0.41 & 0.52 \\ 
        $\sum C_T$ ($\times 10^{-2}$) & 1.02 & 0.92 & 0.72 & 0.62 \\ \hline
        \end{tabular}
}
\end{table}

Table \ref{tbl:1} summarizes the conditions for this verification calculation. 
The rotor speed is 900 RPM. 
The trim conditions are set to satisfy the target total thrust $\sum C_T (= C_T^U + C_T^L)$ and to maintain the balance of rolling moment, pitching moment, and torque. 
The lift offset is assumed to be zero. 
The target thrust is the experimental value for various advance ratios. 
Note that the thrust is not the target value at the rotor trim in the experiment.
In the experiment, the collective pitch angle of the upper rotor was first set to a constant value ($\theta_0 = 8^{\circ}$), and the collective pitch angle of the lower rotor was adjusted to balance the torque. 
Next, the cyclic pitch angles of the upper and lower rotors 
were adjusted to match the target value of the lift offset. 

\begin{figure}
\centering
\begin{tabular}{c}
\begin{minipage}{0.8\columnwidth}
    \includegraphics[width=\columnwidth]{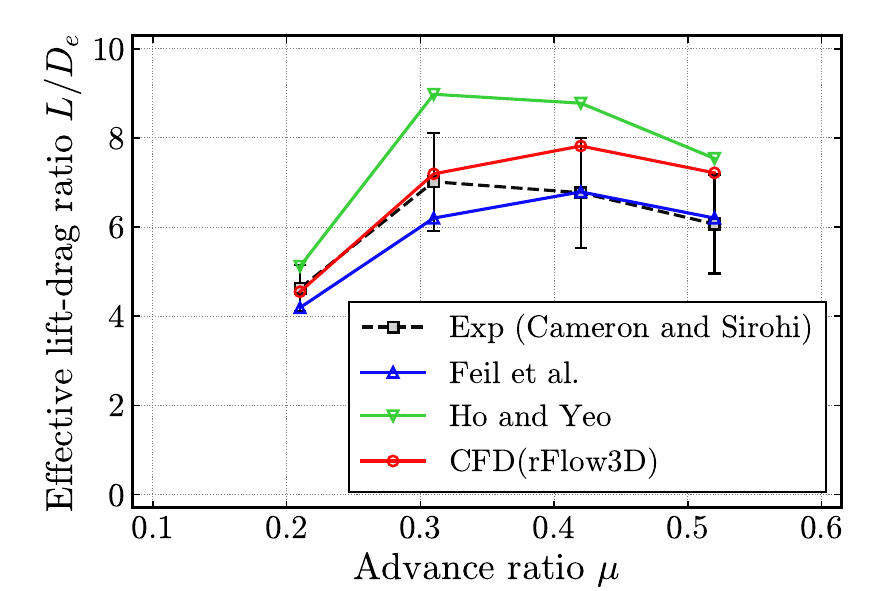}
    \subcaption{ {Effective lift-drag ratio}}
    \label{fig:4_a}
\end{minipage}
\\
\begin{minipage}{0.8\columnwidth}
    \includegraphics[width=\columnwidth]{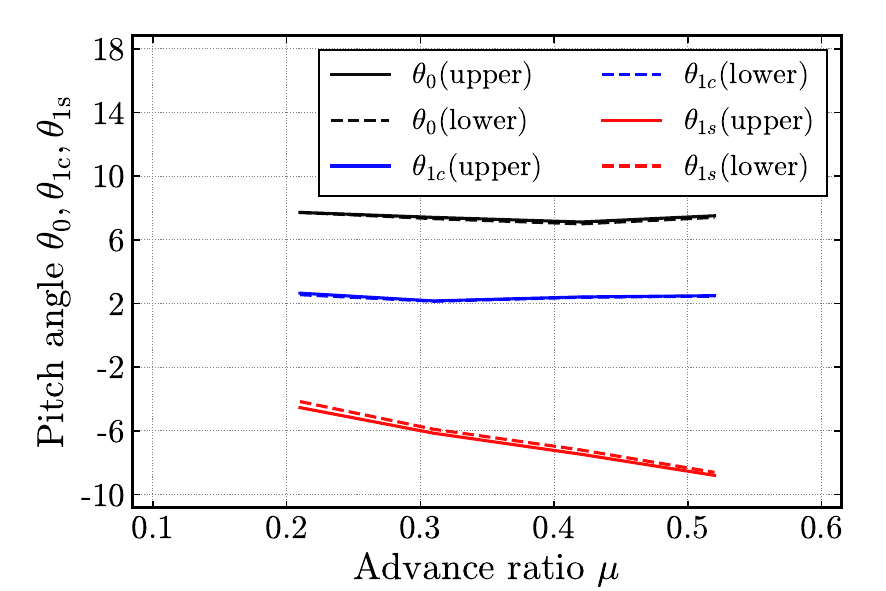}
    \subcaption{Trimmed pitch angles in this calculation}
    \label{fig:4_b}
\end{minipage}
\end{tabular}
\caption{Comparison between various numerical simulations and experiment  {at various advance ratios with $LO =0$.}}
\label{fig:4}
\end{figure}

Figure \ref{fig:4}(a) shows the comparison of various parameters obtained using the rFlow3D with the wind tunnel experiments and the comprehensive analyses. 
In this comparison, the lift-to-effective-drag ratio (effective lift-drag ratio, $L/D_e$) is used as a measure of the aerodynamic performance of the rotor. 
The effective lift-drag ratio is described as
\begin{equation}
    L/D_e = \frac{\sum C_T}{\sum C_P/\mu + \sum C_D},
\label{eq:LbyDe}
\end{equation}
where the rotor power coefficient as $C_P$ and the rotor drag coefficient as $C_D$.
$\sum$ means the total values such as $\sum C_P = C_P^U + C_P^L$.
$\sum C_P$ and $\sum C_D$ are defined as 
\begin{eqnarray}
    \sum C_{P} & = & \frac{\sum P}{\rho_\infty \pi R^2 \left(\Omega R\right)^3}, \label{eq:CP} \\
    \sum C_{D} & = & \frac{\sum D}{\rho_\infty \pi R^2 \left(\Omega R\right)^2}, \label{eq:CD}
\end{eqnarray}
where $P$ represents the rotor power (equivalent to the product of rotor torque $Q$ and rotor angular velocity $\Omega$), and $D$ represents the rotor drag in the forward flight direction.    
The effective lift-drag ratios obtained by this calculation are in good agreement with the experiment and CAMRAD II calculated values. 

Figure \ref{fig:4}(b) shows the trimmed pitch angles from this calculation. 
In the present calculation, the trimmed collective pitch angle for both the upper and lower rotors is approximately 7.5$^{\circ}$, which is consistent with the results of other analyses in which the trimmed collective pitch angle of the lower rotor was about $7^{\circ}$ with the predetermined collective pitch angle of 8$^{\circ}$ for the upper rotor. 
The relation between the longitudinal cyclic pitch angle $\theta_{\rm 1s}$ and the advance ratio $\mu$ also agrees well with Cameron and Sirohi \cite{Cameron2016a} {,} Feil et al. \cite{Feil2019}  {and Ho and Yeo \cite{Ho2020a}}.

Taking into account the difference between the calculated results and the experimental values from the other comprehensive analysis (RCAS) and the error bars in the experiment itself, rFlow3D can be used to properly evaluate the aerodynamic performance of the rotor configuration presented in the following sections. 

\begin{figure}
\centering
  \includegraphics[width=0.8\columnwidth]{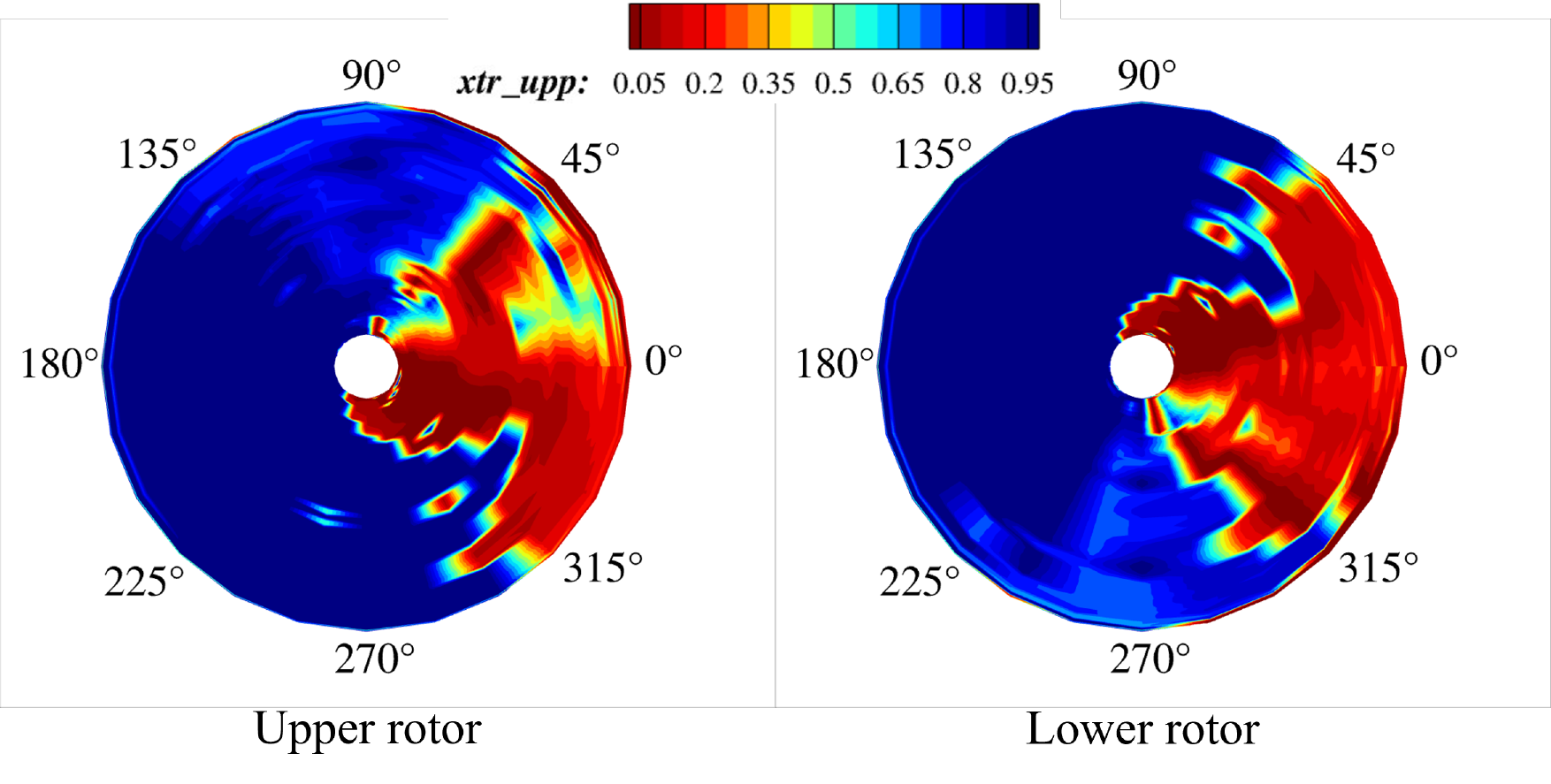}
\caption{Distribution of laminar and turbulent flow distribution on the upper (suction) surface of the blade. Red and blue mean turbulent and laminar flows, respectively.}
\label{fig:5}
\end{figure}

To demonstrate the validity of adopting the transition model, 
the laminar and turbulent distribution of each rotor 
at advance ratio $\mu = 0.53$ is presented in Fig. \ref{fig:5}.
The transition from laminar to turbulent flow occurs at the blade azimuth angle $\Psi \simeq 315^{\circ}$, and returns to the laminar at $\Psi \simeq 45^{\circ}$. 
Therefore, it is reasonable in this calculation to use a computational model that takes into account laminar/turbulent transitions.

\section{Aerodynamic performance}

The aerodynamic performance of the coaxial rotor described in section \ref{sec:2} is analyzed under the conditions of advance ratio $\mu = 0.30 - 0.80$ and lift offset $LO = 0.1 - 0.3$. 
The rotor configuration and computational setup for the analysis
are set as described in previous sections. 
The rotor speed is 900 RPM.
The rotor disk loading $\sum C_T/\sigma$ is assumed to be a constant value ($=7.373 \times 10^{-2}$) in order to equivalently compare the results for all conditions. 
This value is determined with reference to the disk loading of the Sikorsky X2 Technology Demonstrator \cite{Bagai2008}. 
Since the rotor solidity in this calculation is $\sigma =0.1$, 
the total thrust coefficient is $\sum C_T = 7.373 \times 10^{-3}$ for the trim analysis. 
The coefficient of the rolling moment for each rotor is set to a value satisfying the prescribed lift offset $LO$ based on Eq. (\ref{eq:LO}). 
The pitching moment and torque are also set to be balanced across the coaxial rotor. 
The  {inflow angle of freestream refereeing to rotor disk} $i$ is set to 0$^{\circ}$.

\subsection{Trimmed pitch angles}
\label{sec:4_1}

\begin{figure}[t]
\centering
\begin{tabular}{c}
\begin{minipage}{0.8\columnwidth}
    \includegraphics[width=\columnwidth]{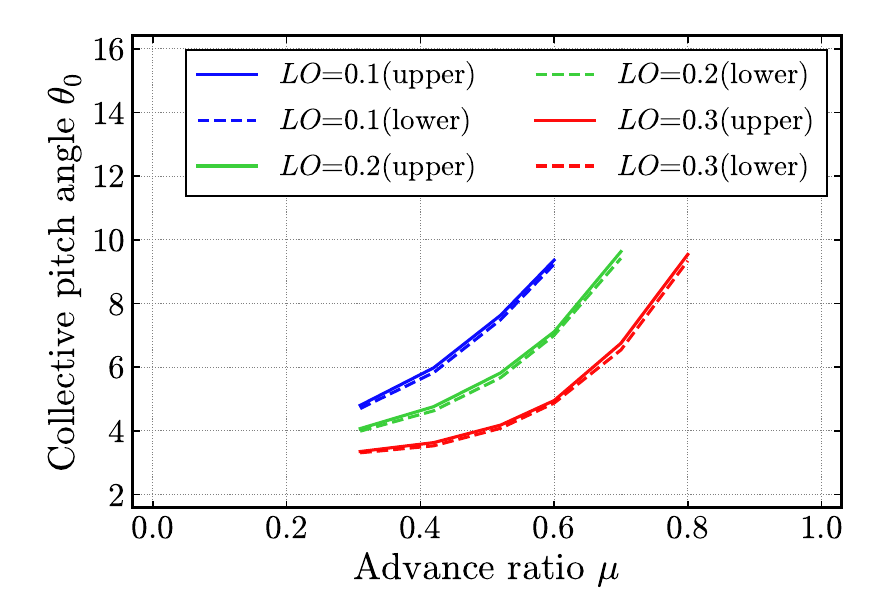}
    \subcaption{Collective pitch angles}
    \label{fig:6_a}
\end{minipage}
\\
\begin{minipage}{0.8\columnwidth}
    \includegraphics[width=\columnwidth]{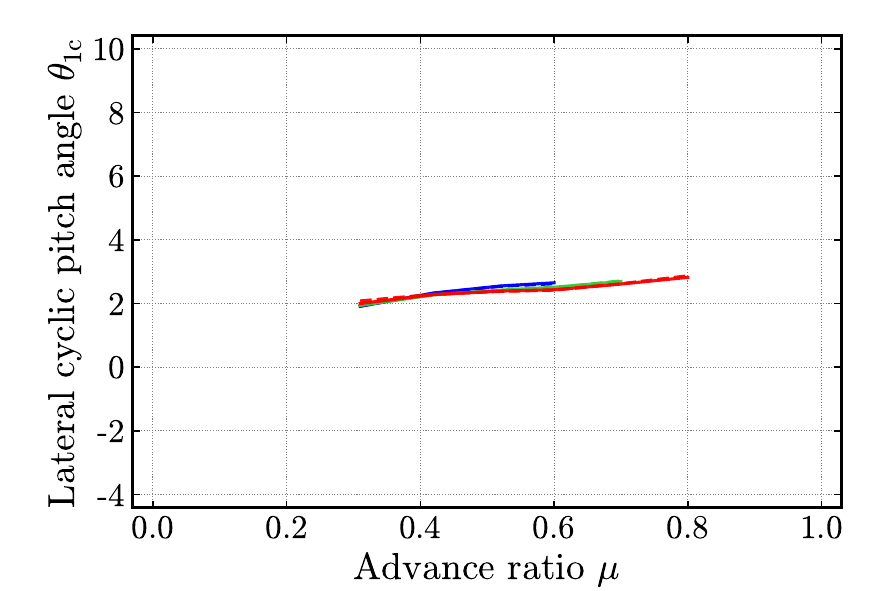}
    \subcaption{Lateral cyclic pitch angles}
    \label{fig:6_b}
\end{minipage}
\\
\begin{minipage}{0.8\columnwidth}
    \includegraphics[width=\columnwidth]{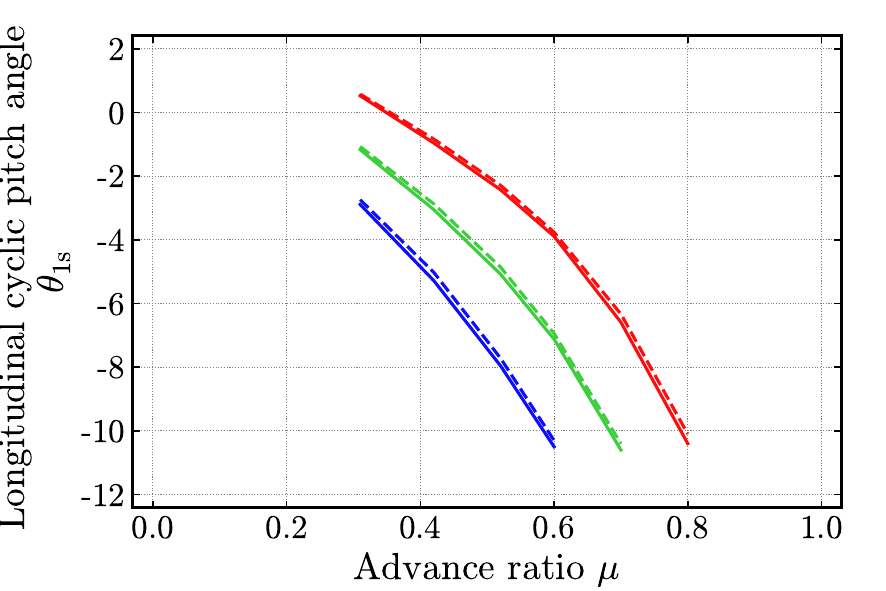}
    \subcaption{Longitudinal cyclic pitch angles}
    \label{fig:6_c}
\end{minipage}
\end{tabular}
\caption{Trimmed pitch angles.}
\label{fig:6}
\end{figure}

Figure \ref{fig:6} shows the pitch angles ($\theta_0$, $\theta_{\rm 1c}$, and $\theta_{\rm 1s}$) obtained from the trim analysis for each condition. 
The results for the upper and lower rotors are represented by solid and broken lines, respectively. 
Note that the trim analysis does not fully converge for the lift offset $LO = 0$ with the advance ratio $\mu$ higher than 0.4. 
As the lift offset $LO$ increases, the upper limit of the advance ratio at which the rotor trim converges increases. 
The lift offset $LO$ must be increased to achieve high-speed flight of a coaxial rotor.

The results show that the collective pitch angle $\theta_0$ increases with an increase in the advance ratio $\mu$ for a constant lift offset. 
The longitudinal cyclic pitch angle $\theta_{\rm 1s}$ is negative and its absolute value also increases with an increase of advance ratio at the constant $LO$. 
The increase in the absolute values of the collective pitch angle $\theta_0$ and the longitudinal cyclic pitch angle $\theta_{\rm 1s}$ is suppressed with an increase in the lift offset $LO$. 
On the other hand, the lateral cyclic pitch angle $\theta_{\rm 1c}$ is hardly affected by the advance ratio $\mu$ and the lift offset $LO$. 
For all conditions, the pitch angles of the upper and lower rotors take almost the same values. 
The dependence of the pitch angle on the advance ratio and lift offset presented in this study is qualitatively similar to the comprehensive analysis by Ho and Yeo \cite{Ho2020b}.

From Eq. (\ref{eq:theta}), the negative longitudinal cyclic pitch angle suppresses the angle of attack on the advancing side of the blade ($\Psi = 90^{\circ}$) and increases the angle of attack on the retreating side of the blade ($\Psi = 270^{\circ}$). 
This control is done to achieve rolling moment balance at a defined $LO$ by reducing lift on the advancing side of the blade and increasing lift on the retreating side of the blade. 
When a lift offset is applied to a coaxial rotor, the constraints on the rolling moment balance are relaxed, i.e., the lift allowed on the advancing side of the blade is increased. 
As the lift offset is increased, the blade pitch angle on the advancing side allowed is also increased, so the negative $\theta_{\rm 1s}$ can be set smaller.  
In particular, it is noteworthy that $\theta_{\rm 1s}$ is almost zero for $\mu = 0.4$ and $LO = 0.3$.

\begin{figure}[t]
\centering
\begin{tabular}{c}
\begin{minipage}{0.8\columnwidth}
    \includegraphics[width=\columnwidth]{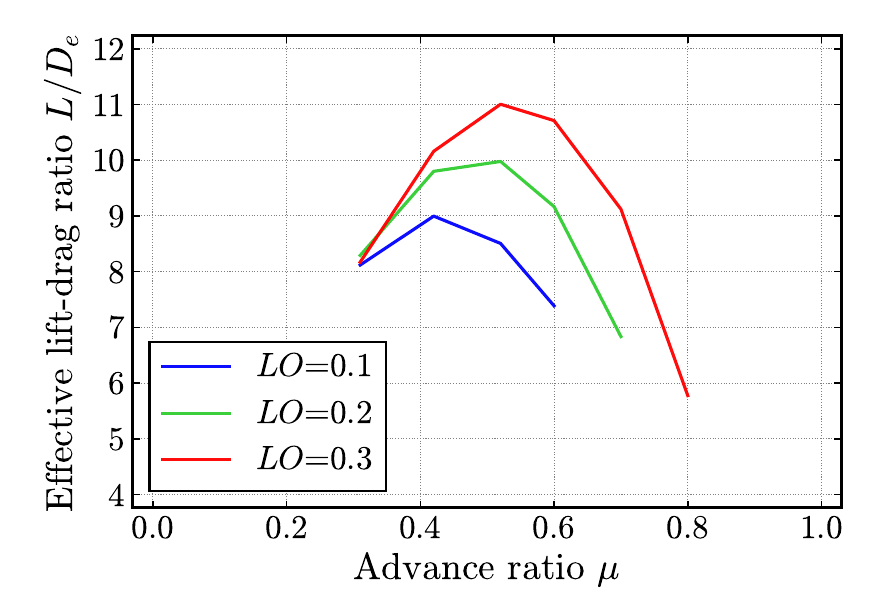}
    \subcaption{Effective lift-drag ratio $L/D_e$}
    \label{fig:7_a}
\end{minipage}
\\
\begin{minipage}{0.8\columnwidth}
    \includegraphics[width=\columnwidth]{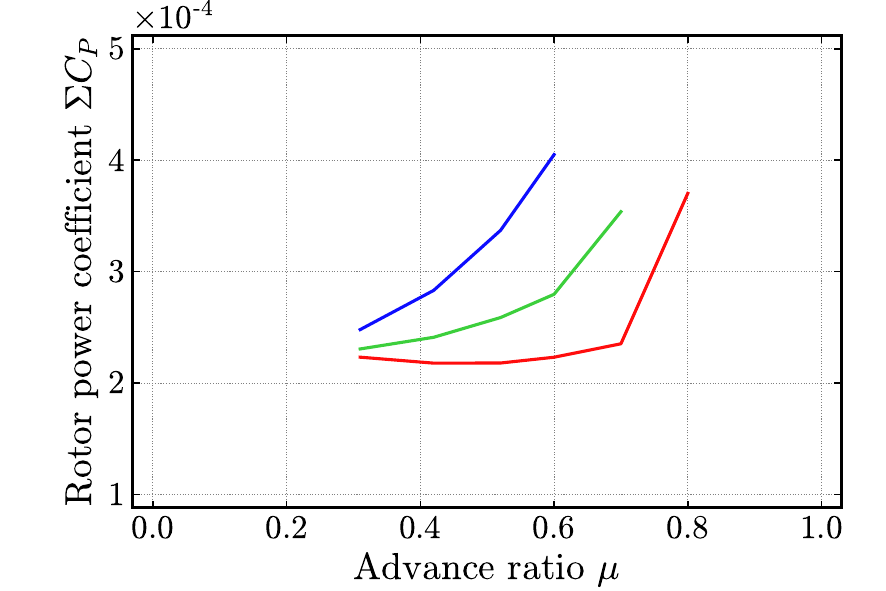}
    \subcaption{Rotor power coefficient $\sum C_P$}
    \label{fig:7_b}
\end{minipage}
\\
\begin{minipage}{0.8\columnwidth}
    \includegraphics[width=\columnwidth]{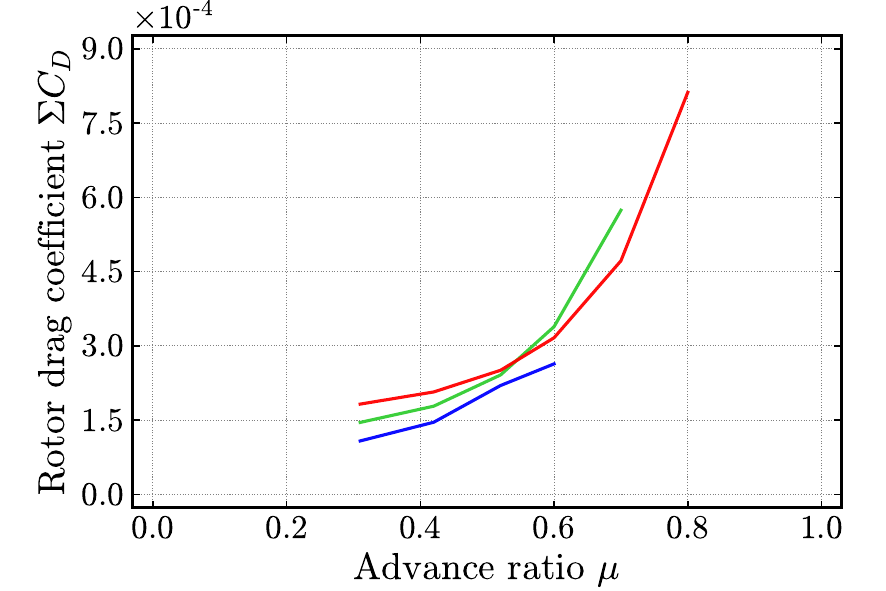}
    \subcaption{Rotor drag coefficient $\sum C_D$}
    \label{fig:7_c}
\end{minipage}
\end{tabular}
\caption{Rotor performance.}
\label{fig:7}
\end{figure}

Under conditions with a high advance ratio ($\mu \ge 0.5$), even with a large lift offset ($LO = 0.3$), the negative value of $\theta_{\rm 1s}$ is larger and the collective pitch angle $\theta_0$ is also larger. 
As the advance ratio increases, the lift on the advancing side of the blade tends to become higher and higher, and it becomes increasingly difficult to ensure lift on the part of the retreating side where the reverse flow occurs. 
Therefore, to reduce the lift difference between the advancing and retreating sides ($\Psi = 90^{\circ}$ and $270^{\circ}$) for maintaining rolling moment balance, the lift at the advancing side is suppressed. 
The collective pitch angle $\theta_0$ is increased to obtain the remainder of the required lift by increasing the lift generated 
in the forward and rear parts ($\Psi = 180^{\circ}$ and $360^{\circ}$) of the rotor.

\subsection{Effective lift-drag ratio}
\label{sec:4_2}

\begin{figure}
\centering
  \includegraphics[width=0.45\columnwidth]{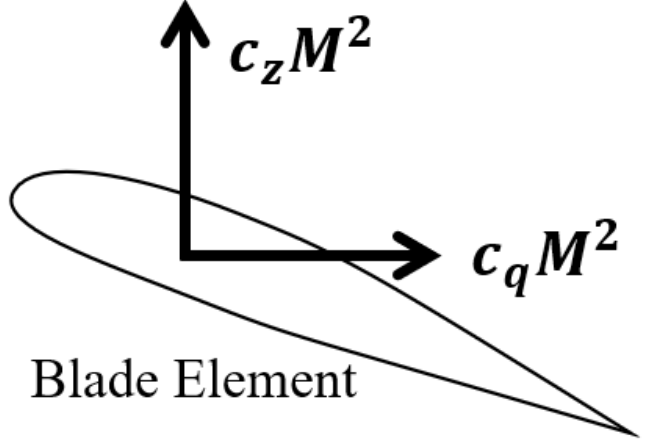}
\caption{Definition of blade element force coefficient.}
\label{fig:8}
\end{figure}

The performance of the coaxial rotor for various advance ratios and lift offsets is summarized in Fig. \ref{fig:7}.  
Figure \ref{fig:7_a} shows the effective lift-drag ratio $L/D_e$ for each condition to clarify the effect of the lift offset on the coaxial rotor.

The effective lift-drag ratio $L/D_e$ is the same as the definition Eq. (\ref{eq:LbyDe}) using the sum of the values of the upper and lower rotors.
Note that the missing parts of the plot correspond to conditions where the trim analysis does not fully converge.

As shown in Figure \ref{fig:7_a}, the effective lift-drag ratio $L/D_e$ has a quadratic form with respect to the advance ratio $\mu$ and reaches a maximum of around $\mu = 0.4$. 
The value of $\mu$ that maximizes $L/D_e$ increases slightly with increasing lift offset.
The maximum effective lift-drag ratio $L/D_e$ increases as the lift-offset $LO$ increases, improving the rotor’s aerodynamic performance. 
However, for $\mu \ge 0.6$, the lift-drag ratio $L/D_e$ decreases significantly with increasing $\mu$, even if the lift offset $LO$ is set to be 0.3.

The values of the total rotor power coefficient $\sum C_P$ and rotor drag coefficient $\sum C_D$ for each advance ratio and lift offset are summarized in Figs. \ref{fig:7_b} and \ref{fig:7_c}, respectively.
Both the rotor power coefficient $\sum C_P$ and the rotor drag coefficient $\sum C_D$ vary with the advance ratio $\mu$ and the lift offset $LO$. 
Since the thrust coefficient $\sum C_T$ is constant for all conditions in this calculation, the variable that causes the effective lift-drag ratio $L/D_e$ to change at each advance ratio is either the rotor power coefficient $\sum C_P$, the rotor drag coefficient $\sum C_D$, or both. 

The rotor power coefficient $\sum C_P$ shown in Fig. \ref{fig:7_b} is lower in the condition with a large lift offset than in the condition with a small lift offset. 
The dependence of $\sum C_P$ on $\mu$ is similar to the relationship between the advance ratio $\mu$ and the collective pitch angle $\theta_0$ shown in Fig. \ref{fig:6_a} except for $\sum C_P$ at $LO = 0.3$ with $\mu = 0.6$ and 0.7. 
It is reasonable to assume that the rotor power decreases as the collective pitch angle decreases because the drag force on the local cross-sectional profile of the entire rotor surface decreases. 
The results suggest that a coaxial rotor can achieve low rotor power flight regardless of the advance ratio $\mu$ by satisfying the lift-offset $LO$ such that the collective pitch angle $\theta_0$ becomes small. 
Note that even with the large lift offset $LO = 0.3$, $\sum C_P$ is significantly large at the very high advance ratio $\mu = 0.8$. 
This point will be discussed with the distribution of sectional drag on the rotor, along with the reason why the shaft power is small for $\mu$ in the range of 0.5 to 0.7 at $LO = 0.3$.

The rotor drag coefficient $\sum C_D$ significantly increases as the advance ratio $\mu$ increases for the high advance ratio $\mu \ge 0.6$, 
From the definition of the effective lift-drag ratio [Eq. (\ref{eq:LbyDe})], the rotor drag coefficient $\sum C_D$ affects the lift-drag ratio in the same order as the rotor power coefficient $\sum C_P$ when the advance ratio $\mu$ is higher than 0.5. 

Ho and Yeo \cite{Ho2020b} showed that setting $LO = 0.3$ improved $L/D_e$ even under conditions $V > 200$ knots ($\mu > 0.6$).
This was partly due to the significant reduction in shaft power as we also show in Fig. \ref{fig:7_b}.  
Another factor was that, unlike our calculations, the increase in drag at $\mu \ge 0.6$ was suppressed.
The cause of this discrepancy will be discussed using the distribution of sectional drag on the rotor.

For a more detailed study of the effect of the lift offset on the rotor power and the drag, we analyze the vertical force coefficient $c_zM^2$ and the sectional drag coefficient $c_qM^2$ for blade element defined in Fig. \ref{fig:8}. 
They are expressed as 
\begin{equation}
    c_zM^2 = \frac{2f_z}{\rho_\infty a_\infty^2 c}, 
    \label{eq:cnm2}
\end{equation}
\begin{equation}
    c_qM^2 = \frac{2f_q}{\rho_\infty a_\infty^2 c},
    \label{eq:ccm2}
\end{equation}
where the vertical force is denoted by $f_z$, the force in the direction of blade movement by $f_q$, speed of sound in  {freestream} by $a_\infty$ and blade chord length by $c$.

\begin{figure}
\centering
  \includegraphics[width=\columnwidth]{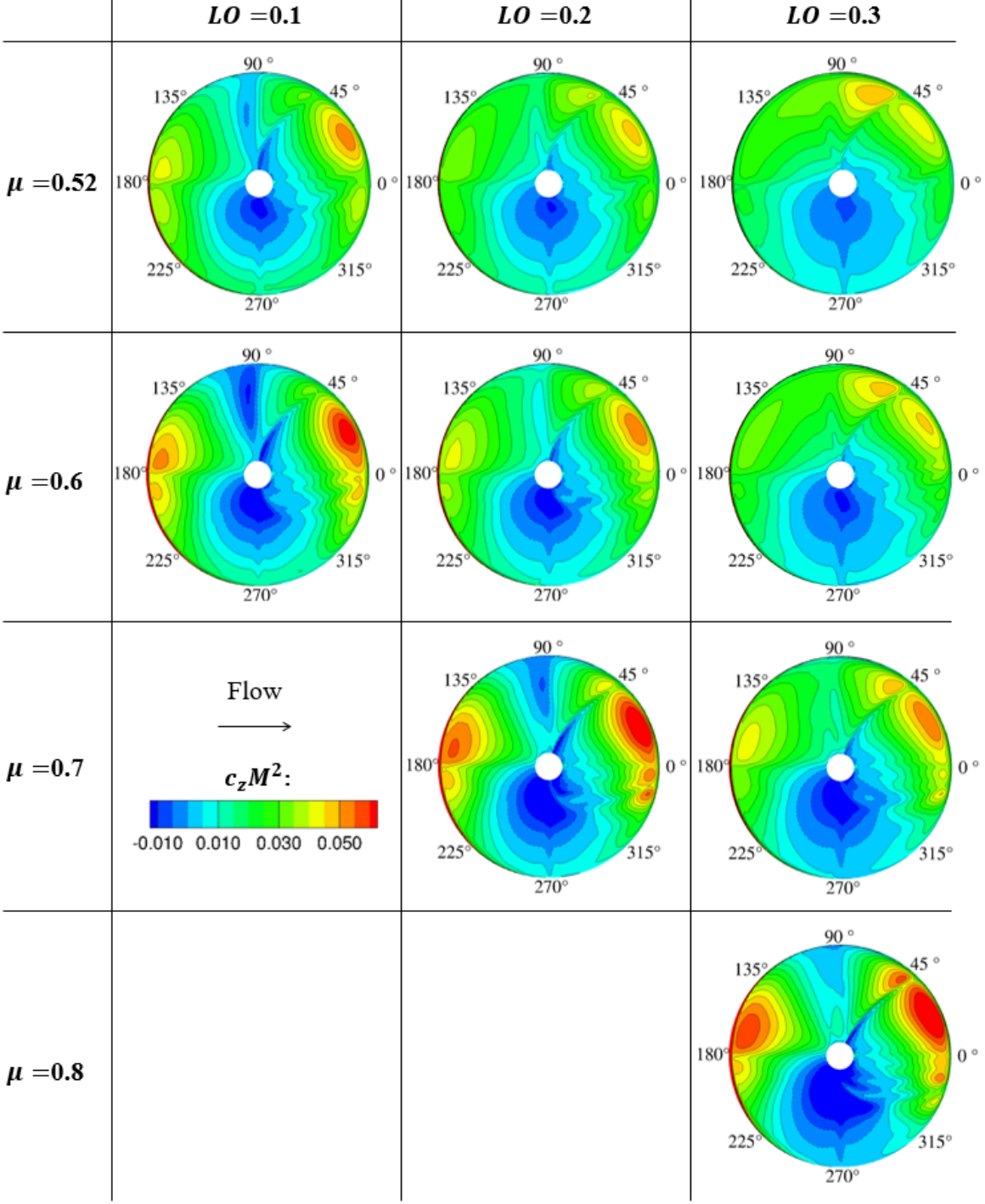}
\caption{Distribution of $c_zM^2$ on the upper rotor disk.}
\label{fig:9}
\end{figure}

\begin{figure}
\centering
  \includegraphics[width=\columnwidth]{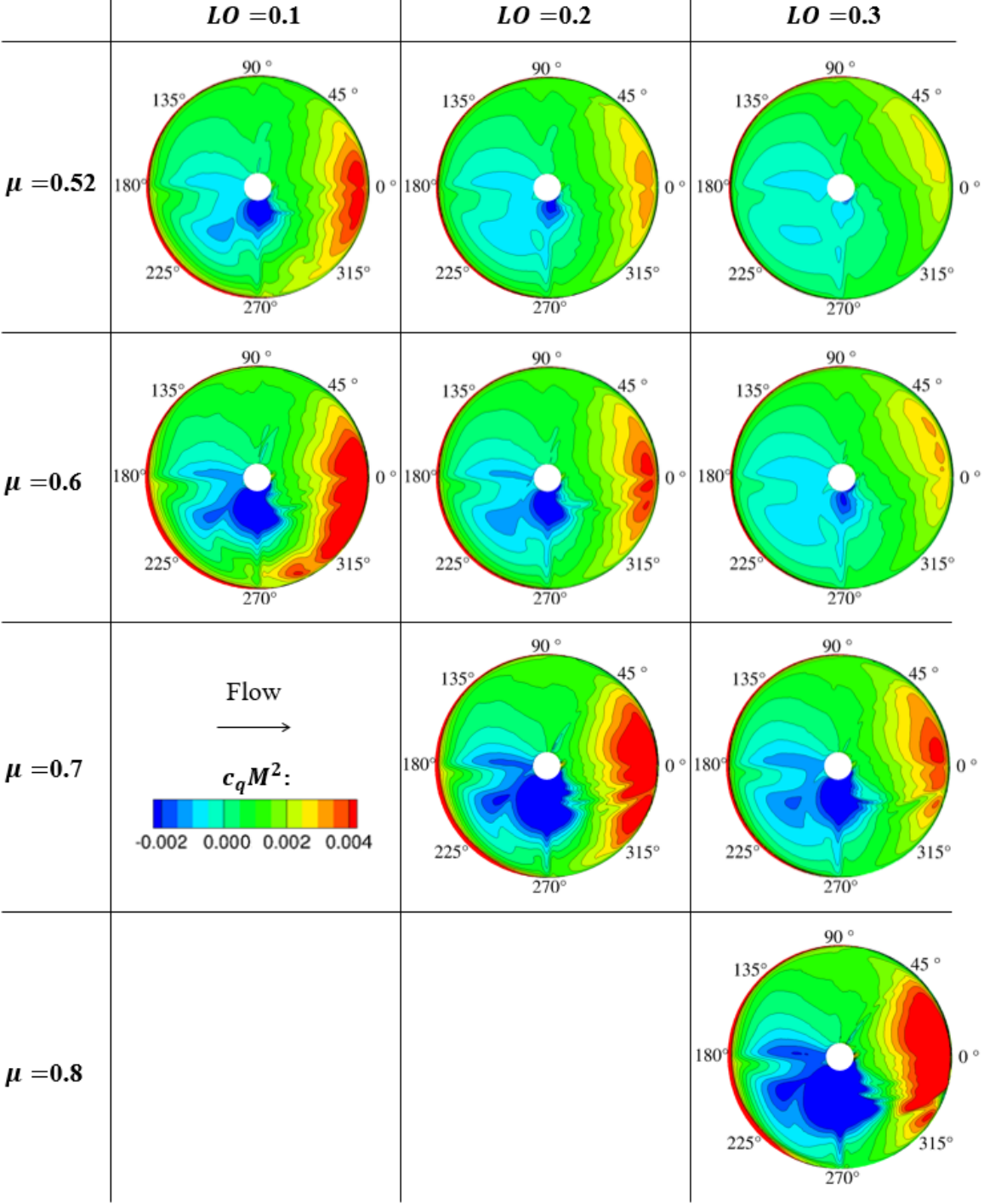}
\caption{Distribution of $c_qM^2$ on the upper rotor disk.}
\label{fig:10}
\end{figure}

Figures \ref{fig:9} and \ref{fig:10} show the distributions of the vertical force (lift) coefficient $c_zM^2$ and the drag coefficient in the direction of blade movement $c_qM^2$ of the upper rotor for various advance ratios and lift offsets, respectively. 
It is clear from Fig. \ref{fig:9} that a significant difference between advance ratio $\mu =0.5$ and 0.8 is found in the distribution of the vertical force coefficient $c_zM^2$ at $LO=0.3$.  
In $\mu = 0.8$, the downforce is observed in large regions at the retreating side of the blade azimuth ($180^{\circ} < \Psi < 360^{\circ}$). 
The region where downforce occurs coincides with the region of reverse flow, whose diameter can be estimated by $\mu R$, 
the product of the advance ratio $\mu$ and the rotor radius $R$. 
The maximum thrust at $\mu \ge 0.7$ is obtained at the blade azimuth angle $\Psi$ around 22.5$^{\circ}$ and 160$^{\circ}$ due to increasing collective pitch angle $\theta_{0}$, as discussed in section \ref{sec:4_1}.

Figure \ref{fig:10} shows that the difference in the sectional drag coefficient $c_qM^2$ distribution for different lift offsets can be seen in this reverse flow region. 
As mentioned in section \ref{sec:4_1}, the blade pitch angle $\theta$ on the retreating side ($180^{\circ} < \Psi < 360^{\circ}$) decreases with increasing lift offset. 
The projected area of the blade relative to the  {freestream} decreases with decreasing blade pitch angle. 
The decrease in the projected area of the blade in  {freestream} owing to the increase in the lift offset reduces the local blade drag, resulting in a decrease in the sectional drag coefficient $c_qM^2$. 
Therefore, for advance ratio $\mu > 0.52$, the rotor drag coefficient $C_D$ is reduced for larger lift offsets $LO$.

At high advance ratios, the negative (blue) value region increases significantly on the retreating side ($180^{\circ} < \Psi < 360^{\circ}$). 
In this region, the blade is pushed in the  {direction of freestream}, which has two effects on aerodynamic performance: a decrease in shaft power and an increase in drag in the uniform flow direction. 
In addition, a very large sectional drag is generated rear of the rotor ($\Psi \simeq 360^{\circ}$), which leads to a significant increase in shaft power. 
 {In forward flight, the rotor disk induces a large downwash at the rear of the rotor \cite{Leishman2006}. 
This downwash increases the induced drag at the rear of the rotor and increases the torque.}

The trimmed pitch angles shown in Fig. \ref{fig:6} indicate that the longitudinal cyclic pitch angle $\theta_{\rm 1s}$ on the retreating side must increase with the advance ratio due to rolling moment balance limitations.
The marked non-uniformity seen in the $c_qM^2$ distribution at the very high advance ratio is due to the nature of this longitudinal cyclic pitch angle.

We use a simple untwisted  {VR-12} blade in this calculation. 
The suppression of increase in drag at $\mu \ge 0.6$ reported by Ho and Yeo \cite{Ho2020b} is probably owing to  {several differences in rotor configuration and operating conditions.  
Ho and Yeo employed a coaxial rotor configuration with four twisted blades per rotor for their analysis.
Their rotor radius was 8.862 m, much larger than ours of 1.016 m.
The rotor speed was reduced with increasing airspeed so that the tip Mach number on the advancing side would not exceed a value of 0.85. 
}

For the improvement of the lift-drag ratio at very high advance ratios, 
it is necessary to optimize the blade shape to reduce not only the collective pitch angle, which contributes to rotor power but also the cyclic pitch angle on the retreating side, which dominates drag in the reverse flow region.

\section{Vibratory  {airloads}}

A CFD analysis using rFlow3D is performed to investigate the effect of lift offset on the thrust fluctuation of an isolated coaxial rotor. 
Thrust fluctuation is a source of vertical oscillation in helicopters flying horizontally.
The computational model and conditions are exactly the same as in the previous section.   
There are three main sources of thrust fluctuation in a coaxial rotor: (1) cyclic variation of vertical force with blade pitch angle $\theta$, (2) pressure field interference due to the blades passing each other, and (3) blade-vortex interaction (BVI). 
The first two sources will be discussed in this study. The discussion on the third source was concisely presented in the authors' previous paper \cite{Hayami2020}. 
 {Note that blade elasticity also contributes to the thrust fluctuation but is neglected in this analysis.}

\begin{figure}
\centering
\begin{tabular}{c}
\begin{minipage}{0.8\columnwidth}
    \includegraphics[width=\columnwidth]{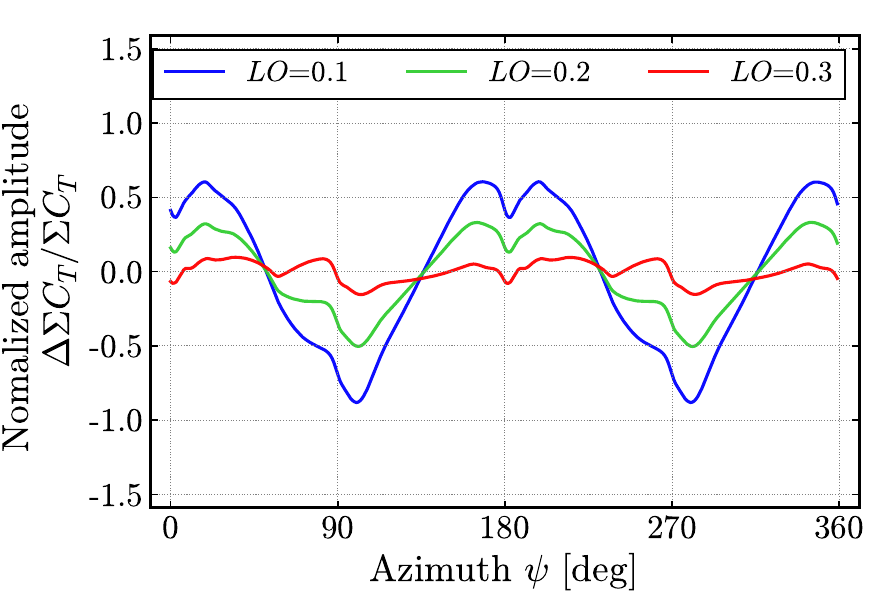}
    \subcaption{Total coefficients}
    \label{fig:11_a}
\end{minipage}
\\
\begin{minipage}{0.8\columnwidth}
    \includegraphics[width=\columnwidth]{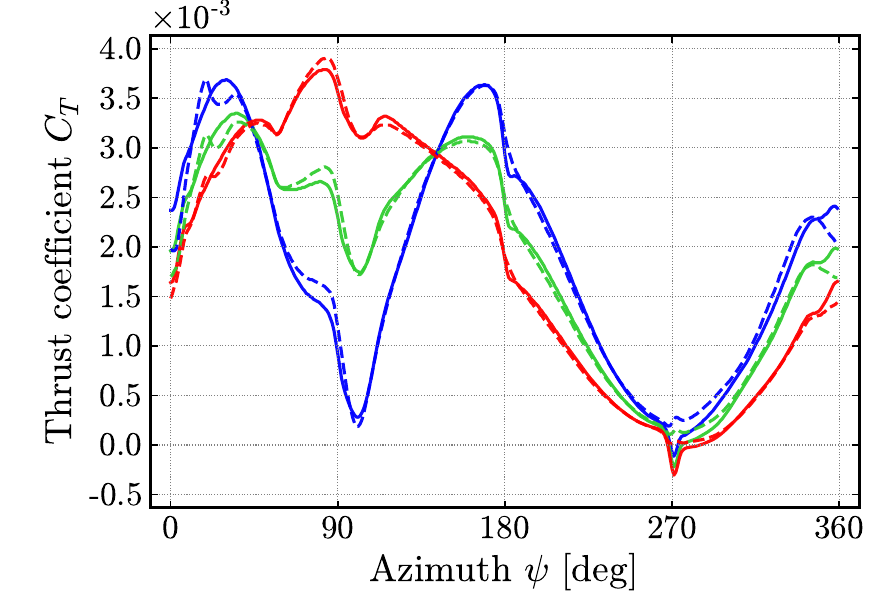}
    \subcaption{Thrust coefficient of each 1 blade of upper and lower rotors: $C_T^U$ (solid) and $C_T^L$ (broken line)}
    \label{fig:11_b}
\end{minipage}
\end{tabular}
\caption{Thrust fluctuation for several values of the lift offset under the advance ratio $\mu=0.52$.}
\label{fig:11}
\end{figure}

Figure \ref{fig:11} shows the variation of the total thrust coefficient of the entire coaxial rotor and the coefficients of single blades in the upper and lower rotors for each azimuth angle $\Psi$ under the advance ratio $\mu$ of 0.52.
For the total thrust coefficient, the dimensionless fluctuation is obtained by subtracting the mean value from the value at each azimuth angle and then dividing it by the mean value.
The total thrust varies periodically with the azimuth angle.
As shown in Fig. \ref{fig:11_a}, the lift offset significantly damps the fluctuation at a constant advance ratio. The peak-to-peak value of the fluctuation in the total thrust is reduced from 1.5 at $LO = 0.1$ to 0.25 at $LO = 0.3$. 

Focusing on the thrust acting on the single blade of each of the upper and lower rotors shown in Fig. \ref{fig:11_b}, the fluctuation on the advancing side ($0^{\circ} < \Psi < 180^{\circ}$) is essentially different with the lift offset. 
For the lift offset of 0.1, a significant drop in the thrust occurs at $\Psi \simeq 90^{\circ}$. 
On the contrary, at $LO = 0.3$, the thrust has a maximum around $\Psi = 90^{\circ}$ with two local minimums around it.
Except for the lower rotor blade with $LO = 0.1$, there is a sharp drop and rise in thrust around $\Psi = 270^{\circ}$.
The azimuth dependence of the thrust acting on a single blade is almost the same for the upper and lower rotors.
However, the blade of the lower rotor has a slightly larger thrust at $\Psi = 90^{\circ}$. 

\begin{figure}
\centering
\begin{tabular}{c}
\begin{minipage}{0.8\columnwidth}
    \includegraphics[width=\columnwidth]{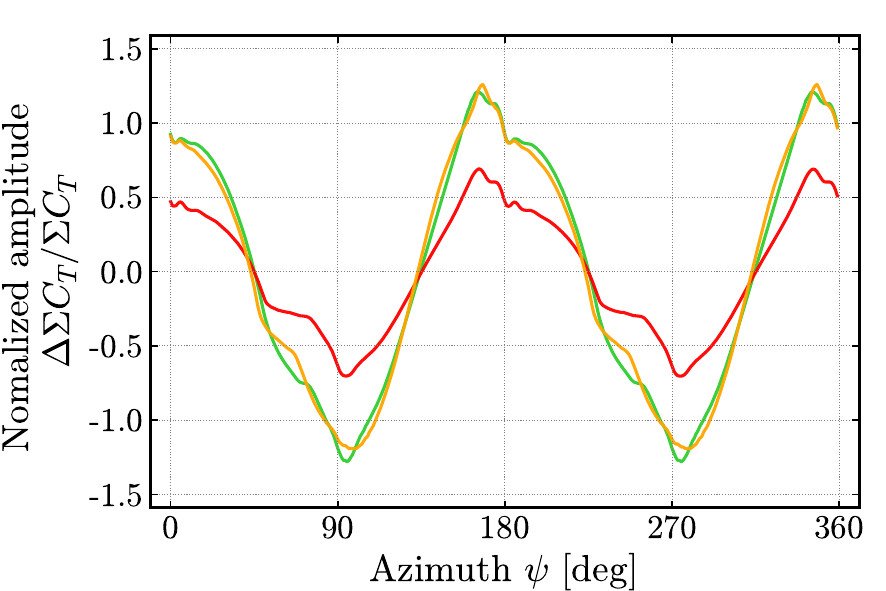}
    \subcaption{Total coefficients}
    \label{fig:12_a}
\end{minipage}
\\
\begin{minipage}{0.8\columnwidth}
    \includegraphics[width=\columnwidth]{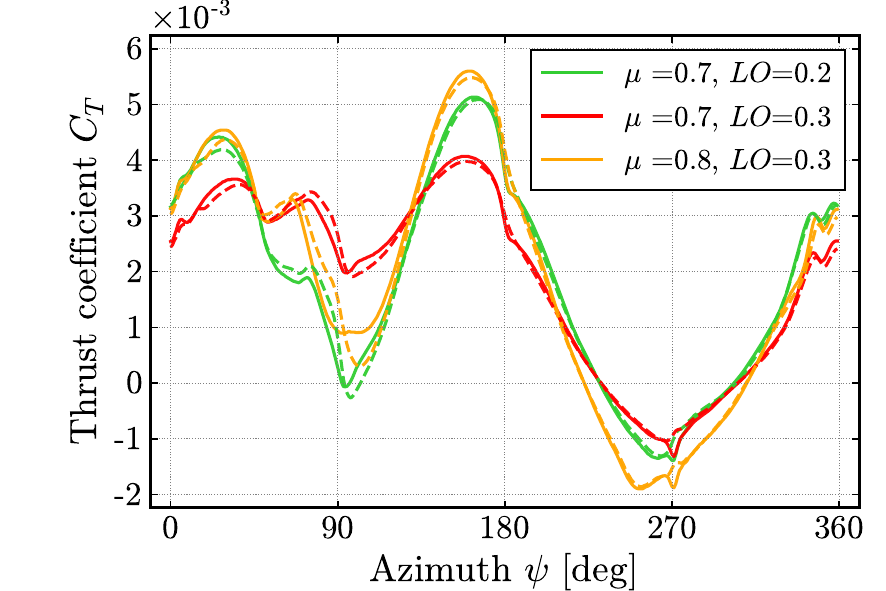}
    \subcaption{Thrust coefficient of each 1 blade of upper and lower rotors: $C_T^U$ (solid) and $C_T^L$ (broken line)}
    \label{fig:12_b}
\end{minipage}
\end{tabular}
\caption{Thrust fluctuations for $\mu=0.7$ and 0.8.}
\label{fig:12}
\end{figure}

Figure \ref{fig:12} shows the thrust fluctuations at the advance ratios $\mu = 0.7$ and 0.8. 
The amplitude of the total thrust fluctuation shown in Fig. \ref{fig:12_a} is significantly large even with the large lift offset. 
The thrusts acting on the single blade of each of the upper and lower rotors shown in Fig. \ref{fig:12_b} indicate that they are negative over a wide range of azimuth angles on the retreating side around $\Psi = 270^{\circ}$. 

Note that the profile of total thrust fluctuation for $\mu = 0.8$ and $LO = 0.3$ is similar to that for $\mu = 0.7$ and $LO = 0.2$. The similarity between the two conditions can also be seen in the distributions of the vertical force and the sectional drag shown in Figs. \ref{fig:9} and \ref{fig:10}.  
However, the thrusts acting on the single blade are different for $\mu = 0.8$ with $LO = 0.3$ and $\mu = 0.7$ with $LO = 0.2$. Moreover, the thrust acting on the upper and lower rotor blades at the same advance ratio and lift offset are slightly different on the advancing side ($0^{\circ} < \Psi < 180^{\circ}$).
The mechanisms of these differences are examined in detail in the subsequent sections.

In the following, the discrete Fourier transform (DFT) is used to analyze these results.
The variation of the thrust coefficient $C_T$ is decomposed into its components with $n$ periods per rotor revolution ($n$/rev), and the amplitude $A_{CT}$ is obtained for each component.

\begin{figure}
\centering
\begin{tabular}{c}
\begin{minipage}{0.8\columnwidth}
    \includegraphics[width=\columnwidth]{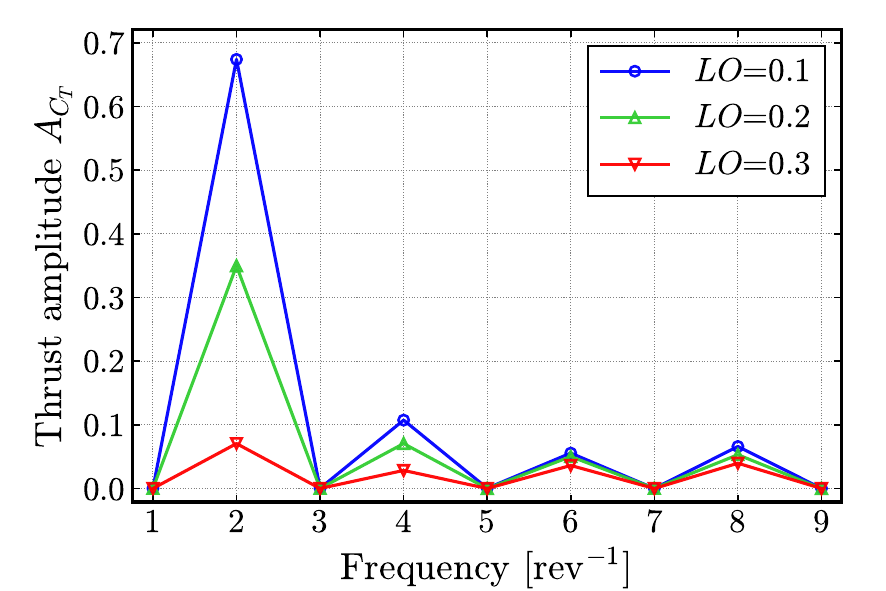}
    \subcaption{$\mu = 0.52$}
    \label{fig:13_a}
\end{minipage}
\\
\begin{minipage}{0.8\columnwidth}
    \includegraphics[width=\columnwidth]{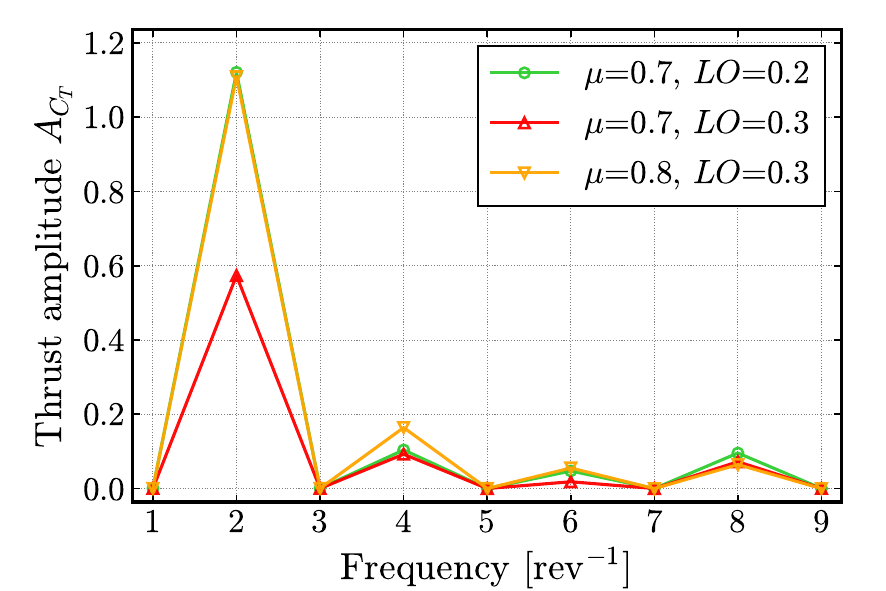}
    \subcaption{$\mu = 0.7$ and 0.8.}
    \label{fig:13_b}
\end{minipage}
\end{tabular}
\caption{Frequency spectra of the total thrust fluctuation.}
\label{fig:13}
\end{figure}

Figure \ref{fig:13} shows the frequency spectra obtained by discrete Fourier transform with the various conditions for advance ratio and lift offset.
Figure \ref{fig:13_a} for $\mu = 0.52$ with lift offset $LO = 0.1$--0.3 indicates that the spectra consist of even-multiplied frequency components.
The amplitude of the 2/rev component dominates the amplitudes of the other $n$/rev components.
Since this trend is similar under a higher advance ratio shown in Fig. \ref{fig:13_b}, 
the effect of the lift offset on the amplitude of the 2/rev component is analyzed first.

\subsection{Thrust variation of 2/rev component}

Feil et al. \cite{Feil2020} analyzed the vibratory loads in a coaxial rotor configuration corresponding to Cameron and Sirohi's experiment as well as our calculations.
They showed the lift offset significantly reduced the 2/rev component of thrust  {variation} in the range of the advance ratio of 0.52 or less.
Here the mechanism of this suppression of thrust fluctuations is analyzed. 
Then the discussion is made whether the suppression is also effective under conditions with a higher advance ratio ($\mu \ge 0.6$).

\begin{figure}
\centering
  \includegraphics[width=0.8\columnwidth]{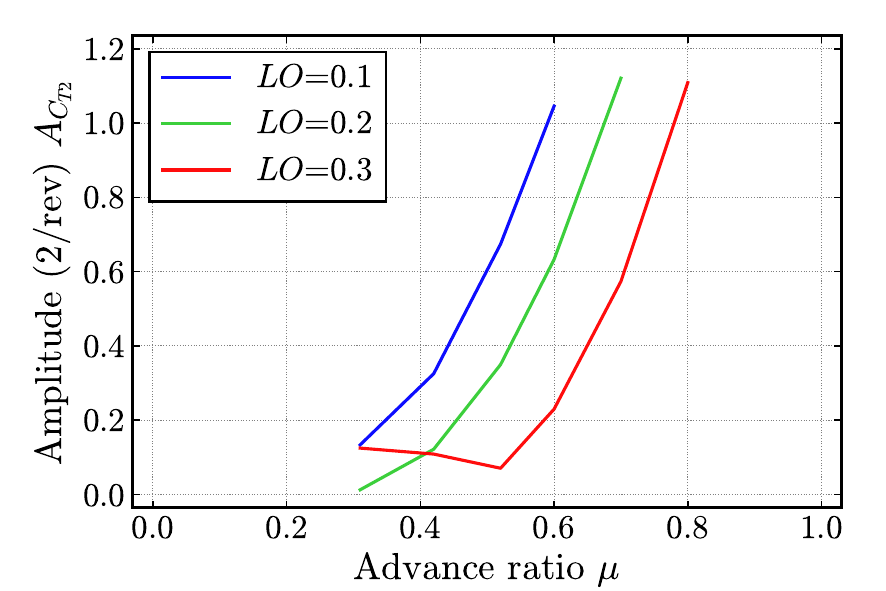}
\caption{The amplitude of the 2/rev  {thrust} components.}
\label{fig:14}
  \includegraphics[width=0.8\columnwidth]{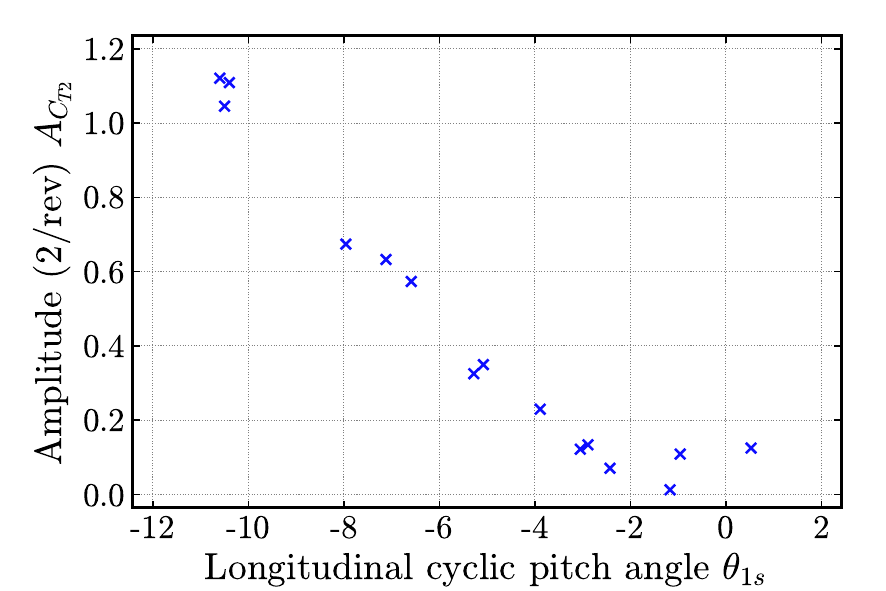}
\caption{Relationship between $\theta_{1s}$ and 2/rev  {thrust variation}.}
\label{fig:15}
\end{figure}

The amplitudes of the 2/rev components for various advance ratios and values of lift offset are compiled in Fig. \ref{fig:14}.
Comparing the amplitudes with different lift offsets $LO$ for a given advance ratio $\mu$ shows that the amplitudes decrease significantly as the lift offset is increased.
In particular, for $LO = 0.3$, the 2/rev component is less than 0.1 in the range $\mu < 0.6$.
For $\mu \ge 0.6$, however, the amplitude increases rapidly with increasing advance ratio, even with a large lift offset ($LO = 0.3$).

In this calculation, one rotor consists of two blades. 
The thrust at any given azimuth angle $\Psi$ is the sum of the lift forces of the blade located at $\Psi$ and the blade located at $\Psi + 180^{\circ}$.
That is, as can be seen in Fig. \ref{fig:11_a}, the thrust fluctuation at azimuth angles $\Psi$ of around  $0^{\circ}$ and $180^{\circ}$ (or around $90^{\circ}$ and $270^{\circ}$) is almost the same value, and its magnitude is the sum of the lift forces at front and rear of the rotor (or on the advancing and retreating sides of the rotor).

\begin{figure}
\centering
\begin{tabular}{c}
\begin{minipage}{0.8\columnwidth}
    \includegraphics[width=\columnwidth]{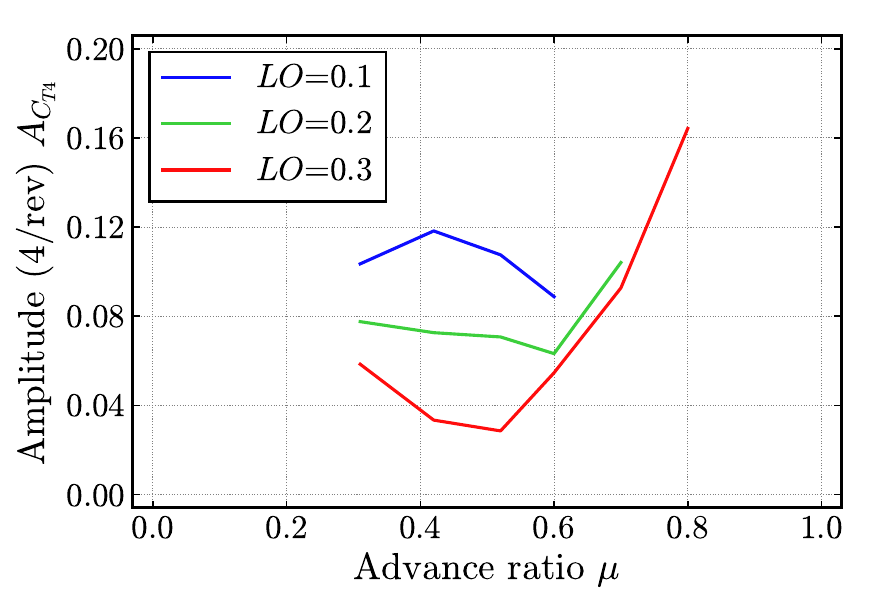}
    \subcaption{total}
    \label{fig:16_a}
\end{minipage}
\\
\begin{minipage}{0.8\columnwidth}
    \includegraphics[width=\columnwidth]{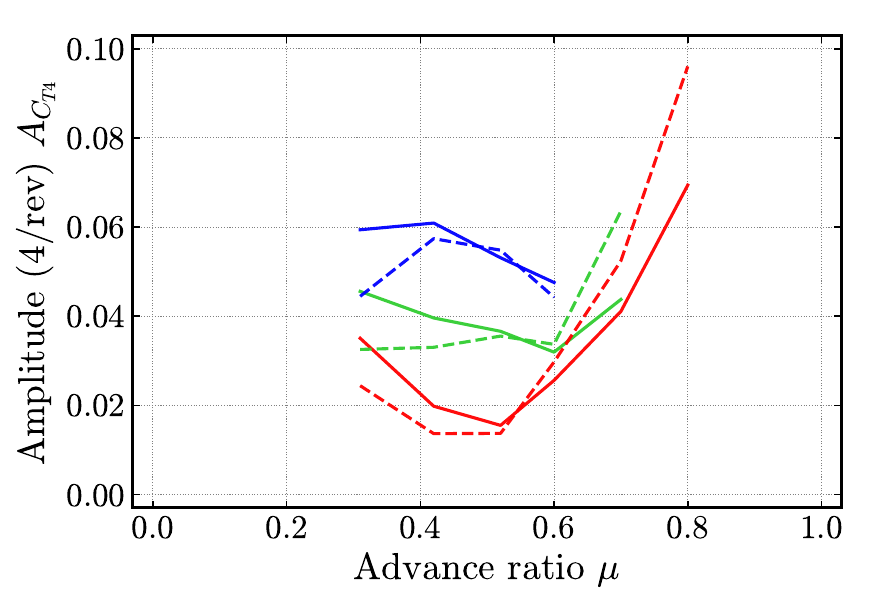}
    \subcaption{upper \& lower}
    \label{fig:16_b}
\end{minipage}
\end{tabular}
\caption{The amplitude of the 4/rev  {thrust} components. {\sout{.}}}
\label{fig:16}
\end{figure}

The large amplitude of the 2/rev component is due to the large lift generated near the front and the rear of the rotor in the forward direction at the high advance ratio with the small lift offset (see Fig. \ref{fig:9}). 
This results in a large thrust force at azimuth angles $\Psi = 0^{\circ}$ and $180^{\circ}$.
The thrust is small at azimuths $\Psi = 90^{\circ}$ and 270$^{\circ}$ because the lift is almost vanished on the advancing and retreating sides of the rotor.

Figure \ref{fig:15} shows the relationship between the longitudinal cyclic pitch angle $\theta_{\rm 1s}$ and the amplitude of the 2/rev component for all 15 results obtained in this calculation.
The amplitude of the 2/rev component can be correlated by the longitudinal cyclic pitch angle $\theta_{\rm 1s}$. 
Among the rotor controlling angles ($\theta_0$, $\theta_{\rm 1c}$, and $\theta_{\rm 1s}$), the longitudinal cyclic pitch angle $\theta_{\rm 1s}$ is the factor varies the magnitude of the pitch angle (i.e. lift) on the rotor's advancing and retreating sides.
Reducing the longitudinal cyclic pitch angle to zero, i.e., increasing the lift of the advancing side blade, is effective not only in improving the lift-drag ratio but also in reducing the 2/rev component fluctuation.

Note that the 2/rev component of  {vibratory aerodynamics loards} may be reduced by adjusting the first blade crossing angle of two rotors \cite{Feil2020, Ho2020a}. 
This point will be discussed in the future.

\begin{figure*}[t]
\centering
  \includegraphics[width=0.8\textwidth]{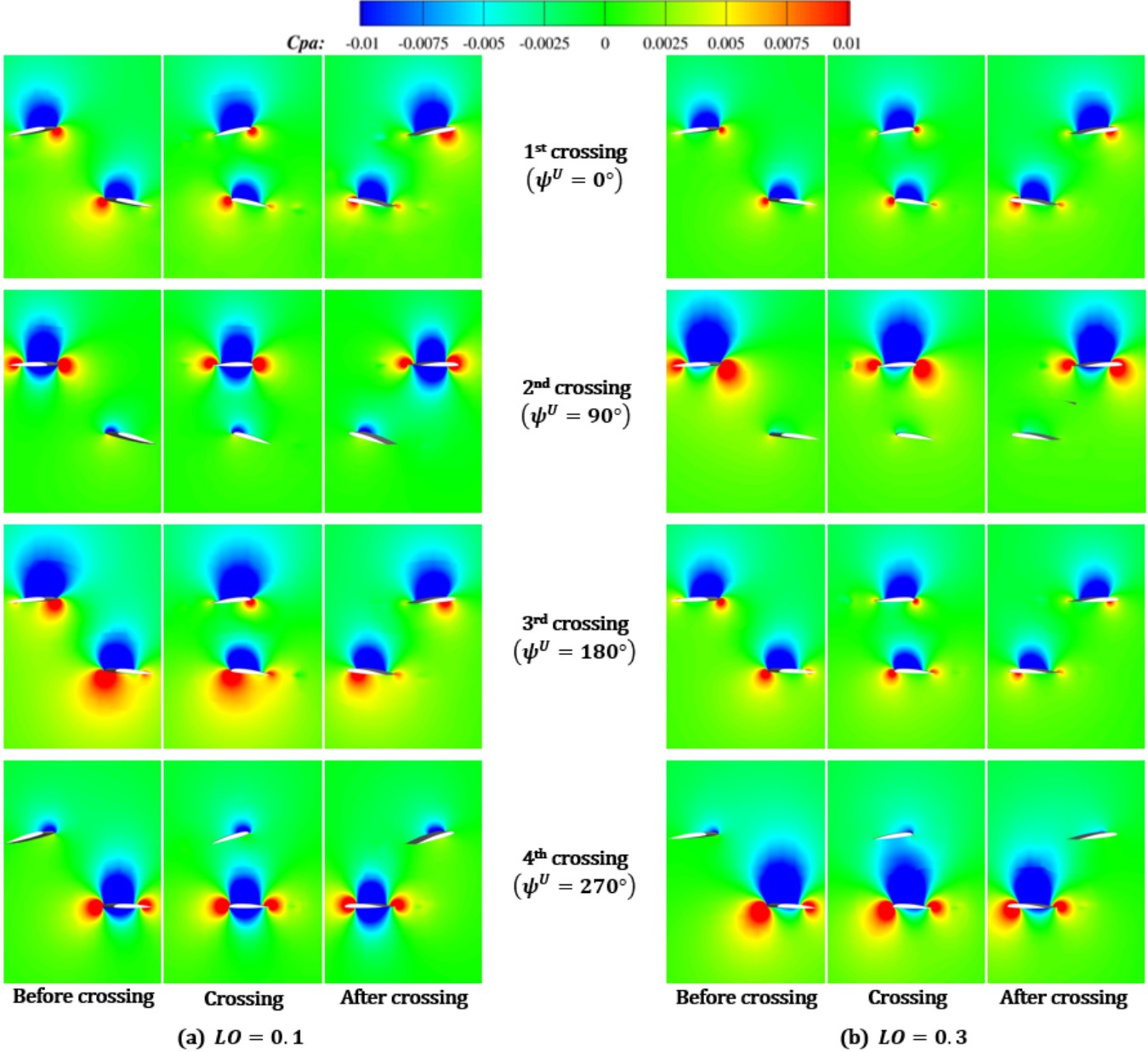}
\caption{Contours of pressure coefficient around the two blades at their crossover events for $\mu=0.52$.}
\label{fig:17}
\end{figure*}

\subsection{Thrust variation of 4/rev component}

The 4/rev component is analyzed from the present calculations given various lift offsets at advance ratios higher than 0.5.
The major source of this component is due to blade crossing events. 
Passe et al. \cite{Passe2015} showed impulsive air loads corresponding to blade passage in an 8-bladed coaxial rotor (4 blades per rotor) in their calculations with a single lift offset given for each advance ratio in the range $\mu \le 0.41$, according to the X2 Technology Demonstrator (X2TD) experiment. 

\begin{figure}
\centering
  \includegraphics[width=0.8\columnwidth]{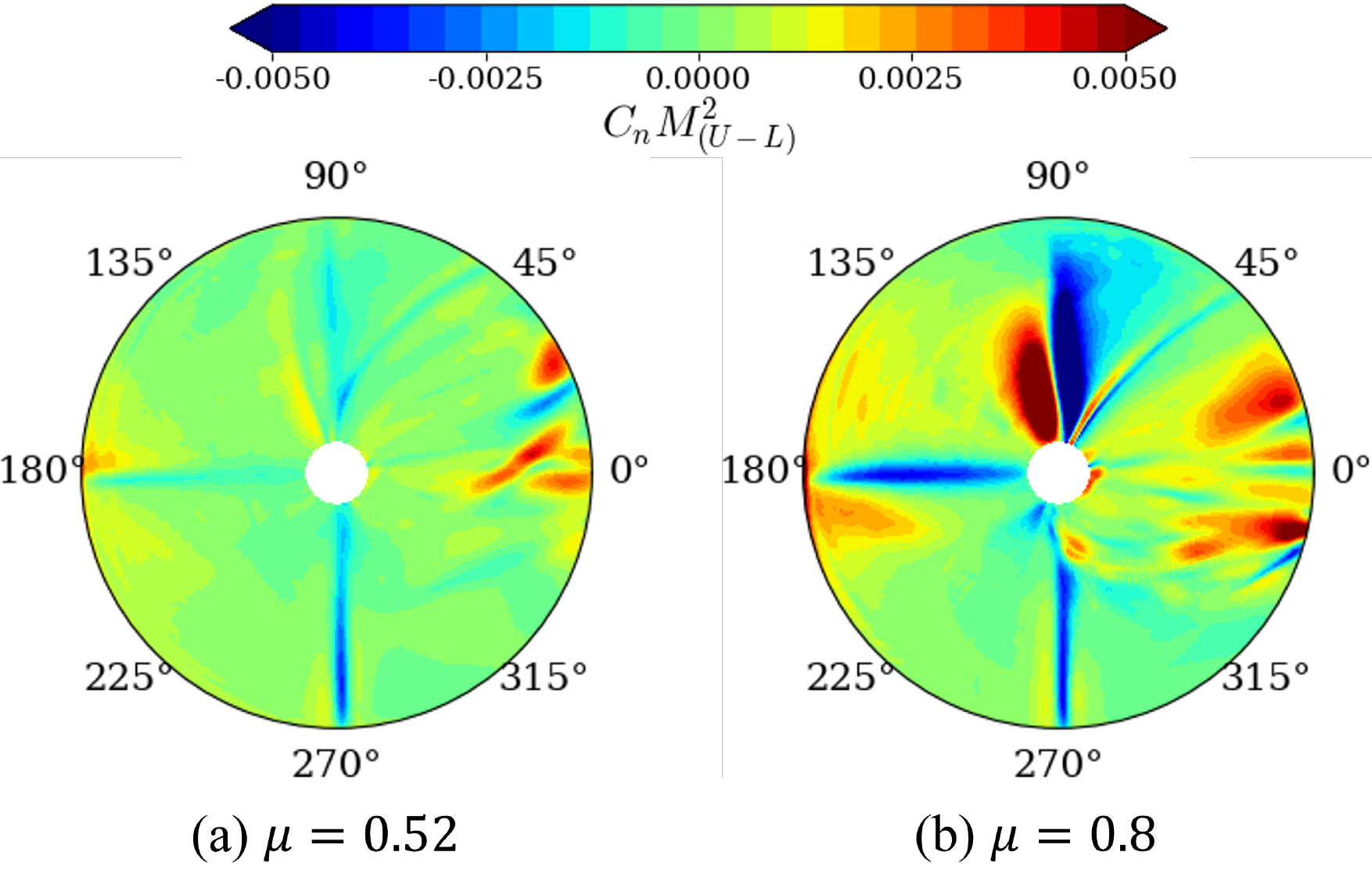}
\caption{Distribution of $C_nM^2$ difference between upper and lower rotor disk.}
\label{fig:18}
\end{figure}

Figure \ref{fig:16} shows the amplitude of the 4/rev component of the thrust fluctuation. 
The 4/rev component of the total thrust shown in Fig. \ref{fig:16_a} indicates that the lift offset reduces the amplitude at the advance ratio $\mu = 0.52$. 
 {
As the advance ratio increases, the amplitude increases even with a large lift offset.}
Notice that the amplitude for $\mu = 0.8$ with $LO = 0.3$ in total thrust fluctuation is larger than for $\mu = 0.7$ with $LO = 0.2$, where the amplitudes of 2/rev component are the same with each other (see Fig. \ref{fig:14}). 
This implies that the 4/rev component is not just a harmonic of the 2/rev component.
Figure \ref{fig:16_b} shows that the amplitudes of the thrust fluctuations for the upper and lower rotors are nearly equal for $\mu = 0.52$, whereas the amplitude of the lower rotor is larger at $\mu > 0.6$. 
 {The cause of this difference is discussed later using Fig. \ref{fig:18}.
}

Returning to the azimuth dependence of the total thrust (Figs. \ref{fig:11_a} and \ref{fig:12_a}), the thrust drops  {\sout{sharply}} at azimuth angles $\Psi = 0^{\circ}, 90^{\circ}, 180^{\circ}$, and 270$^{\circ}$. 
These azimuth angles are when the upper and lower rotor blades pass through each other.
When the blades pass through each other, the pressure field on the pressure surface of the upper rotor blade interferes with the pressure field on the suction surface of the lower rotor blade.
As a result of the interference, the difference between the positive and negative pressure on each blade becomes so small that the lift force decreases. This reduction in lift causes fluctuations in thrust due to the blades crossing each other.

Figure \ref{fig:17} shows the change in contour maps of the pressure $p$ as the blades pass through each other for the advance ratio $\mu = 0.52$. 
These pressure distributions are for cross sections through $r = 0.75R$ perpendicular to the $r$-axis for each phase angle $\Psi^U$. 
Figure \ref{fig:17}(a) corresponds to the case $LO = 0.1$ and Fig. \ref{fig:17}(b) to the case $LO = 0.3$.
The effect of lift offset on the mutual interference of pressure fields when blades pass through each other is discussed using this figure. 
The crossing events are labeled as from 1st to 4th crossing in the order of azimuth angle of the upper rotor ($\Psi = 0^{\circ}$, 90$^{\circ}$, 180$^{\circ}$ and 270$^{\circ}$).
The contours are displayed using the pressure coefficient, 
\begin{equation}
    C_{pa} = \frac{2(p- p_\infty)}{\rho_\infty a_\infty^2}.  
\label{eq:cpa}
\end{equation}
where $p_\infty$ denotes the static pressure in  {freestream}. 

Focusing on the 1st crossing, the high-pressure region of the upper blade for $LO = 0.1$ (Fig. \ref{fig:17}(a)) shrinks as the low-pressure region of the lower blade approaches.
This high-pressure region recovers after the blades pass through each other. 
In contrast, the pressure coefficient for $LO = 0.3$ (Fig. \ref{fig:17}(b)) has a small variation range for both the upper and lower blades. 
The small pressure variation around the blades at $LO = 0.3$ is due to the small rotor thrust required at this crossing angle ($\Psi = 0^{\circ}$) (see Fig. \ref{fig:9}).
The above trend is also observed for the third crossing of the upper and lower rotor blades at azimuth angle $\Psi = 180^{\circ}$, where the pressure field for $LO = 0.1$ is slightly different from the 1st crossing.

In the case of 2nd crossing, the upper blade is located on the advancing side of the rotor, while the lower blade is located on the retreating side of the rotor.
At $LO = 0.1$ shown in Fig. \ref{fig:17}(a), the pitch angle $\theta$ of the lower rotor blade is so large that the low-pressure region above its suction surface enhances the low pressure below the upper rotor blade after the crossing.
In contrast, for $LO =0.3$ shown in Fig. \ref{fig:17}(b), the pitch angle $\theta$ of the lower blade is small. Both the upper and lower rotor blades do not show any change in the pressure field during the blade crossing event.
The pressure interference weakens as the lift offset $LO$ increases.

At the 4th crossing, the upper rotor blade is located on the retreating side of the rotor, while the lower rotor blade is located on the advancing side of the rotor. 
Comparing Figs. (a) and (b) in Fig. \ref{fig:17}, the low-pressure area above the lower rotor blade is larger for $LO =0.3$ than for $LO =0.1$, which also has a larger effect on the upper rotor blade pressure field. This is the cause of the steep thrust drop seen at azimuth angle $\Psi = 270^{\circ}$ shown in Fig \ref{fig:11_b}.

In summary, lift offset suppresses thrust fluctuations when the rotor blades cross in the front and rear locations. It also suppresses the fluctuations when the upper blades are on the advancing side. In contrast, lift offset slightly amplifies thrust fluctuations when the upper blade is on the retreating side.
In general, as the pitch angle increases, the pressure distribution around the blade becomes more pronounced. Therefore, keeping both the collective and cyclic pitch angles small is effective in weakening pressure interference during blade crossing.

Finally, we discuss the 4/rev component at $\mu = 0.8$. 
Figure \ref{fig:18} shows the difference in the vertical force coefficient $C_nM^2$ between the upper and lower rotors at the advance ratios $\mu = 0.52$ and 0.8 with the lift offset $LO = 0.3$. 
Note that it represents the difference between the upper and lower rotors at the same azimuth angle.
The red color means the $C_nM^2$ of the upper rotor is larger than that of the lower rotor, and the blue color means the opposite. 
The sharp blue line at $\Psi = 270^{\circ}$ corresponds to the thrust drop of the upper rotor blade at the 4th blade crossing as discussed in  Fig. \ref{fig:16}. 
 {Compared to $\mu = 0.52$, for $\mu = 0.8$, $C_nM^2$ of the lower rotor is substantially larger (blue-colored regions in Fig. \ref{fig:18}) at $\Psi = 90^{\circ}$ and $180^{\circ}$ than the value of the upper rotor.}
Complex distribution of mixed red and blue is seen around $\Psi = 0^{\circ}$.
Since this is the rear of the rotor disk, the complex distribution may be attributed to the flow disturbance generated by the front blades.
 {These features lead} to an increase in the 4/rev component at $\mu = 0.8$ as well as the difference in amplitudes for the upper and lower rotors.

\section{Conclusions}

The aerodynamic performance of an isolated coaxial rotor in high-speed forward flight is analyzed using computational fluid dynamics (CFD) calculations.
Lift-drag ratios and thrust fluctuations are investigated in detail under feasible lift-offset conditions (0.3 or less) for a range of advance ratios from 0.3 to 0.8. 

A four-blade system (two blades per rotor) is used in the analysis because detailed wind tunnel experimental data \cite{Cameron2016a} and comprehensive analysis results \cite{Feil2019, Ho2020a} are available for this system.
Comparisons with these previous studies confirm that our numerical calculations are sufficiently accurate.

Then calculations are performed with various lift offsets for high-speed flight conditions with advance ratios higher than 0.5. 
The principal conclusions are summarized as follows: 
\begin{enumerate}
    \item Applying the lift offset  {at a constant advance ratio} improves the lift-to-effective drag ratio (lift-drag ratio) and reduces thrust fluctuations. 
 {
This application is particularly useful for keeping both the collective pitch angle and the longitudinal cyclic pitch angle small for all advance ratios. 
A small collective pitch angle improves the lift-drag ratio through the reduction of rotor power. 
A small longitudinal cyclic pitch angle reduces the thrust fluctuation.}
    \item In the case where the advance ratio exceeds 0.6, the lift-drag ratio drops significantly even if the lift offset is 0.3. 
    The thrust fluctuation also increases with such a high advance ratio. 
 {
For example, the lift-drag ratio $L/D_e = 5.8$  at $\mu = 0.8$ and $LO = 0.3$ is 40\% less than $L/D_e = 11$ at $\mu = 0.5$ and $LO = 0.3$.  
The amplitude of the 2/rev component of the thrust variation $A_{C_{T2}}$ at $\mu = 0.8$ and $LO = 0.3$ is larger than the target value of the thrust. 
}
\item The degradation of aerodynamic performance and  {vibratory aerodynamic loads} is closely related to the pitch angle control to compensate for the reduction in thrust on the retreating side due to the increased reverse flow region. 
 {
For example, the collective pitch angle $\theta_0 = 9^{\circ}$ at $\mu = 0.8$ and $LO = 0.3$ is twice that at $\mu = 0.6$ and $LO = 0.3$.  
The longitudinal cyclic pitch angle $\theta_{1s} = 10^{\circ}$ at $\mu = 0.8$ and $LO = 0.3$ is 2.5 times larger than $\theta_{1s}$ at $\mu = 0.6$ and $LO = 0.3$. 
}
    \item The results suggest that it is effective to reduce the collective and longitudinal cyclic pitch angles for the improvement of the aerodynamic performance of coaxial rotors.
\end{enumerate}

\section*{Acknowledgement}

This research was supported by Grant-in-Aid for Scientific Research (B), Grant No. JP22H01396, from the Japan Society for the Promotion of Science (JSPS KAKENHI).
The calculations were performed using the Japan Aerospace Exploration Agency's supercomputer system (JSS3).

%\appendix

 \bibliographystyle{elsarticle-num} 
 \bibliography{cas-refs}





% \end{thebibliography}
\end{document}